\documentclass[a4paper,10pt]{article} %
\usepackage{latexsym}
\usepackage{amssymb} %
\usepackage{amsmath} %
\usepackage{amsthm} %
\usepackage{amscd} %
\title{
Joint Extension of States of Subsystems for a CAR System}
\author{Huzihiro Araki \thanks{Mailing address: 
Research Institute for Mathematical Sciences, Kyoto 
 University.  Kitashirakawa-Oiwakecho, Sakyoku, Kyoto 606-8502
 Japan} 
\, and Hajime Moriya}
\date{}
\begin{document}
\maketitle
\theoremstyle{plain}%
\newtheorem{thm}{Theorem}
\newtheorem{df}[thm]{Definition}%
\newtheorem{lem}{Lemma}[section]%
\newtheorem{cor}[lem]{Corollary}
\newtheorem{pro}[thm]{Proposition}%
\theoremstyle{remark}%
\newtheorem*{rem}{{Remark}}
\numberwithin{equation}{section}
\newcommand{\R}{{\mathbb{R}}}%
\newcommand{\Z}{{\mathbb{Z}}}%
\newcommand{\Com} {{\mathbb{C}}}%
\newcommand{\NN}{{\mathbb{N}}}%
\newcommand{\qedb}{\hbox{\rule[-2pt]{3pt}{6pt}}}%
\newcommand{\Al}{{\cal A}}%
\newcommand{\Bl}{{\cal B}}%
\newcommand{\Cl}{{\cal C}}%
\newcommand{\DAl}{{\mathfrak{ A} }}%
\newcommand{\DBl}{{\mathfrak{ B} }}%
\newcommand{\DCl}{{\mathfrak{ C} }}%
\newcommand{\DMl}{{\mathfrak{ M} }}%
\newcommand{\DMlpm}{ \DMl_{\pm}}%
\newcommand{\DMlp}{ \DMl_{+}}%
\newcommand{\DMlm}{ \DMl_{-}}%
\newcommand{\nonum}{\nonumber}%
\newcommand{\Mat}{{\mathrm  M}_{2}(\C)}%
\newcommand{\Matfour}{{\mathrm M}_{4}(\C)}%
\newcommand{\aicr}{a_i^{\ast}}%
\newcommand{\ai}{a_i}%
\newcommand{\akcr}{a_k^{\ast}}%
\newcommand{\ak}{a_k}%
\newcommand{\aicrhat}{\hat{a_i^{\ast}}}%
\newcommand{\aihat}{\hat{a_i}}%
\newcommand{\aicrcheck}{\check{a_i^{\ast}}}%
\newcommand{\aicheck}{\check{a_i}}%
\newcommand{\ajcr}{a_j^{\ast}}%
\newcommand{\aj}{a_j}%
\newcommand{\uone}{u_{1}}%
\newcommand{\utwo}{u_{2}}%
\newcommand{\uonehat}{\widehat{u}_1}%
\newcommand{\ui}{u_{i}}%
\newcommand{\Uone}{U_{1}}%
\newcommand{\Utwo}{U_{2}}%
\newcommand{\Uf}{U_{1}}%
\newcommand{\Ui}{U_{i}}%
\newcommand{\Uimh}{U_{i-1}}%
\newcommand{\Uzero}{U_{0}}%
\newcommand{\Util}{\tilde{U}}%
\newcommand{\Ufb}{\breve{U}_{1}}%
\newcommand{\Uib}{\breve{U}_{i}}%
\newcommand{\Uzerob}{\breve{U}_{0}}%
\newcommand{\Wi}{W_{i}}%
\newcommand{\Wf}{W_{1}}%
\newcommand{\Wzero}{W_{0}}%
\newcommand{\vI}{v_{\I}}%
\newcommand{\ji}{j_{i}}
\newcommand{\uiz}{u^{(i)}_{0}}%
\newcommand{\uih}{u^{(i)}_{1}}%
\newcommand{\uif}{u^{(i)}_{2}}%
\newcommand{\uim}{u^{(i)}_{3}}%
\newcommand{\uiji}{u^{(i)}_{\ji}}%
\newcommand{\Uji}{U_{\{ {\ji} \} } }%
\newcommand{\cstar}{{\bf C}^{\ast}}%
\newcommand{\wstar}{{\bf W}^{\ast}}%
\newcommand{\I}{{\mathrm{I}}}%
\newcommand{\J}{{\mathrm{J}}}%
\newcommand{\K}{{\mathrm{K}}}%
\newcommand{\LL}{{\mathrm{L}}}%
%
\newcommand{\Io}{\I_{0}}%
\newcommand{\If}{\I_{1}}%
\newcommand{\Is}{\I_{2}}%
\newcommand{\Ii}{\I_{i}}%
\newcommand{\Ij}{\I_{j}}%
\newcommand{\Ik}{\I_{k}}%
\newcommand{\Il}{\I_{l}}%
\newcommand{\In}{\I_{n}}%
%
\newcommand{\Iik}{\I_{\ik}}%
\newcommand{\Iione}{\I_{\ione}}%
\newcommand{\Iis}{\I_{\is}}%
\newcommand{\Ifup}{\cup_{i \ge 1}\Ii}%
\newcommand{\Isup}{\cup_{i \ge 2}\Ii}%
\newcommand{\vrho}{\varrho}%
\newcommand{\vrhopri}{\varrho^{\prime}}%
\newcommand{\psif}{\psi_{1}}%
\newcommand{\psis}{\psi_{2}}%
\newcommand{\psii}{\psi_{i}}%
\newcommand{\psil}{\psi_{l}}%
\newcommand{\psij}{\psi_{j}}%
\newcommand{\chii}{\chi_{i}}%
\newcommand{\chik}{\chi_{\kappa}}%
\newcommand{\chikl}{\chi_{kl}}%
\newcommand{\ome}{\omega}%
\newcommand{\omeA}{\ome_{\Al}}%
\newcommand{\omeB}{\ome_{\Bl}}%
\newcommand{\omeC}{\ome_{\Cl}}%
\newcommand{\sig}{\sigma}%
\newcommand{\vareps}{\varepsilon}%
\newcommand{\eps}{\epsilon}%
\newcommand{\lam}{\lambda}%
\newcommand{\lami}{\lam_{i}}%
\newcommand{\lamj}{\lam_{j}}%
\newcommand{\lamk}{\lam_{k}}%
\newcommand{\mul}{\mu_{l}}%
%
\newcommand{\lamvppsi}{\lam(\vp,\,\psi)}%
\newcommand{\lamvps}{\lam(\vps)}%
\newcommand{\sigj}{\sig(j)}%
\newcommand{\sigji}{\sig(\{\ji\})}%
\newcommand{\sigp}{\sig^{\prime}}%
\newcommand{\pvpf}{p(\vpf)}%
\newcommand{\pvps}{p(\vps)}%
%
%
\newcommand{\xip}{\xi_{+}}%
\newcommand{\xim}{\xi_{-}}%
\newcommand{\xipm}{\xi_{\pm}}%
\newcommand{\xp}{x_{+}}%
\newcommand{\xm}{x_{-}}%
\newcommand{\xpm}{x_{\pm}}%
%
\newcommand{\vp}{\varphi}%
\newcommand{\vpf}{\vp_{1}}%
\newcommand{\pis}{\pi_{2}}%
\newcommand{\pispri}{\pis^{\prime}}%
\newcommand{\pipri}{\pi^{\prime}}%
\newcommand{\vpfp}{\vp_{1+}}%
\newcommand{\vpfm}{\vp_{1-}}%
\newcommand{\vpfpm}{\vp_{1\pm}}%
\newcommand{\vps}{\vp_{2}}%
\newcommand{\vpsp}{\vp_{2+}}%
\newcommand{\vpsm}{\vp_{2-}}%
\newcommand{\vpspm}{\vp_{2\pm}}%
\newcommand{\vpal}{\vp_{\alpha}}%
\newcommand{\vpbe}{\vp_{\beta}}%
\newcommand{\vpbeta}{\vp_{\beta}}%
\newcommand{\vpfal}{\vp_{1\alpha}}%
\newcommand{\vpfbe}{\vp_{1\beta}}%
\newcommand{\vpsal}{\vp_{2\alpha}}%
\newcommand{\vpsbe}{\vp_{2\beta}}%
\newcommand{\vpb}{\overline{\varphi}}%
\newcommand{\vpfb}{\overline{\vpf}}%
\newcommand{\vpsb}{\overline{\vps}}%
\newcommand{\vphat}{\hat{\vp}}%
\newcommand{\vpshat}{\widehat{\vp}_{2}}%
\newcommand{\vpfhat}{\widehat{\vp}_{1}}%
\newcommand{\pishat}{\widehat{\pi}_{2}}%
\newcommand{\pifhat}{\widehat{\pi}_{1}}%
\newcommand{\vpshatt}{\vpshat \Theta}%
\newcommand{\vpfhatt}{{\widehat{\vpf}}\Theta}%
%
%
%
\newcommand{\hvps}{\widehat{\vp}_{2}}%
\newcommand{\hvpst}{\widehat{\vp}_{2}\Theta}%
\newcommand{\vppri}{\vp^{\prime}}%
\newcommand{\vpfpri}{\vpf^{\prime}}%
\newcommand{\vpspri}{\vps^{\prime}}%
\newcommand{\vpsprit}{\vps^{\prime}\Theta}%
\newcommand{\vpo}{\vp_{0}}%
\newcommand{\vpi}{\vp_{i}}%
\newcommand{\vpj}{\vp_{j}}%
\newcommand{\vpk}{\vp_{k}}%
\newcommand{\vpik}{\vp_{\ik}}%
\newcommand{\vpione}{\vp_{\ione}}%
\newcommand{\vpis}{\vp_{\is}}%
\newcommand{\pii}{\pi_{i}}%
\newcommand{\pio}{\pi_{0}}%
\newcommand{\piiup}{\pi^{i}}%
\newcommand{\pioup}{\pi^{0}}%
\newcommand{\piib}{\breve{\pi}_{i}}%
\newcommand{\piob}{\breve{\pi}_{0}}%
%
\newcommand{\piitil}{\tilde{\pi_{i}}}%
%
\newcommand{\Ikup}{\I^{k}}%
\newcommand{\vpkup}{\vp^{k}}%
\newcommand{\vpkmoup}{\vp^{k-1}}%
\newcommand{\vpfup}{\vp^{1}}%
\newcommand{\vpsup}{\vp^{2}}%
%
\newcommand{\AlI}{{\cal A}({\I})}%
\newcommand{\AlJ}{{\cal A}({\J})}%
\newcommand{\AlIhat}{\hat{\cal A}({\I})}%
\newcommand{\AlJhat}{\hat{\cal A}({\J})}%
\newcommand{\AlIfhat}{\hat{\cal A}({\If})}%
\newcommand{\AlIshat}{\hat{\cal A}({\Is})}%
\newcommand{\AlIcheck}{\check{\cal A}({\I})}%
\newcommand{\AlJcheck}{\check{\cal A}({\J})}%
\newcommand{\Ali}{{\cal A}(\{i\})}%
%
%
\newcommand{\AlIo}{{\cal A}({\Io})}%
\newcommand{\AlIf}{{\cal A}({\If})}%
\newcommand{\AlIs}{{\cal A}({\Is})}%
\newcommand{\AlJf}{{\cal A}({\Jf})}%
\newcommand{\AlJs}{{\cal A}({\Js})}%
\newcommand{\AlIi}{{\cal A}({\Ii})}%
\newcommand{\AlIj}{{\cal A}({\Ij})}%
\newcommand{\AlIik}{{\cal A}({\Iik})}%
\newcommand{\AlIione}{{\cal A}({\Iione})}%
\newcommand{\AlIis}{{\cal A}({\Iis})}%
\newcommand{\AlcupIi}{\Al({\cup_{i}\Ii})}%
\newcommand{\AlIot}{{\cal A}({\If\cup \Itwo})}%
\newcommand{\AlIfs}{{\cal A}({\If\cup \Is})}%
\newcommand{\Alp}{{\Al}_{+}}%
\newcommand{\Alm}{{\Al}_{-}}%
\newcommand{\Alpm}{{\Al}_{\pm}}%
\newcommand{\AlIp}{{\AlI}_{+}}%
\newcommand{\AlIm}{{\AlI}_{-}}%
\newcommand{\AlIpm}{{\AlI}_{\pm}}%
%
\newcommand{\Alip}{{\Ali}_{+}}%
\newcommand{\Alim}{{\Ali}_{-}}%
\newcommand{\Alipm}{{\Ali}_{\pm}}%
\newcommand{\AlJp}{{\AlJ}_{+}}%
\newcommand{\AlJm}{{\AlJ}_{-}}%
\newcommand{\AlJpm}{{\AlJ}_{\pm}}%
\newcommand{\AlIip}{{\AlIi}_{+}}%
\newcommand{\AlIim}{{\AlIi}_{-}}%
\newcommand{\AlIipm}{{\AlIi}_{\pm}}%
\newcommand{\AlIfp}{{\AlIf}_{+}}%
\newcommand{\AlIfm}{{\AlIf}_{-}}%
\newcommand{\AlIfpm}{{\AlIf}_{\pm}}
\newcommand{\AlIsp}{{\AlIs}_{+}}%
\newcommand{\AlIsm}{{\AlIs}_{-}}%
\newcommand{\AlIspm}{{\AlIs}_{\pm}}%
\newcommand{\AlK}{{\cal A}({\K})}%
\newcommand{\AlL}{{\cal A}({\LL})}%
\newcommand{\AlJuI}{{\cal A}({\J}\cup {\I})}%
\newcommand{\AlIuJ}{{\cal A}({\I}\cup {\J})}%
\newcommand{\AlIfus}{{\cal A}({\If}\cup {\Is})}
\newcommand{\id}{{\mathbf{1} } }
\newcommand{\Aone}{A_{1}}%
\newcommand{\Atwo}{A_{2}}%
\newcommand{\ione}{i_{1}}%
\newcommand{\is}{i_{2}}%
\newcommand{\ik}{i_{k}}%
\newcommand{\ij}{i_{j}}%
%
%
%
\newcommand{\Af}{A_{1}}%
\newcommand{\As}{A_{2}}%
\newcommand{\Ao}{A_{0}}%
\newcommand{\All}{A_{l}}%
\newcommand{\Aj}{A_{j}}%
\newcommand{\Ai}{A_{i}}%
\newcommand{\Ak}{A_{k}}%
\newcommand{\Ap}{A_{+}}%
\newcommand{\Am}{A_{-}}%
\newcommand{\Apm}{A_{\pm}}%
\newcommand{\vpt}{{\vp\Theta}}%
\newcommand{\vpft}{{\vpf\Theta}}%
\newcommand{\vpst}{{\vps\Theta}}%
\newcommand{\pivp}{\pi_{\vp}}%
\newcommand{\pivptil}{\tilde{\pivp}}%
\newcommand{\pivpt}{\pi_{\vp\Theta}}%
\newcommand{\pivps}{\pi_{\vps}}%
\newcommand{\pivpsp}{\pi_{\vpsp}}
\newcommand{\pivpspri}{\pi_{\vpspri}}%
\newcommand{\pivpst}{\pi_{\vpst}}%
\newcommand{\pivpstil}{\pi_{\vpstil}}
\newcommand{\vpstil}{{\widetilde{\vp}}_{2}}%
\newcommand{\vpstilt}{{\widetilde{\vp}}_{2}\Theta}%
%
\newcommand{\pihvps}{\pi_{\hvps}}%
\newcommand{\pihvpst}{\pi_{\hvpst}}%
\newcommand{\pipsi}{\pi_{\psi}}%
\newcommand{\pitau}{\pi_{\tau}}%
\newcommand{\pivpf}{\pi_{\vpf}}%
\newcommand{\pivpft}{\pi_{\vpf\Theta}}%
\newcommand{\Pvppsi}{P(\vp,\,\psi)}%
\newcommand{\Pvpvpt}{P(\vp,\,\vp\Theta)}%
\newcommand{\Pvpfvpft}{P(\vpf,\,\vpf\Theta)}%
\newcommand{\omeh}{\ome^{h}}%
\newcommand{\Hil}{{\cal H}}%
\newcommand{\Hild}{\widehat{\Hil}}%
\newcommand{\Hilds}{\widehat{\Hil}_{2}}%
\newcommand{\Hili}{\Hil_{i}}%
\newcommand{\Hilo}{\Hil_{0}}%
\newcommand{\Hilj}{\Hil_{j}}%
\newcommand{\Hilf}{\Hil_{1}}%
\newcommand{\Hils}{\Hil_{2}}%
\newcommand{\Hilvp}{{\cal H}_{\vp}}%
\newcommand{\Hilvpf}{{\cal H}_{\vpf}}%
\newcommand{\Hilvps}{{\cal H}_{\vps}}%
\newcommand{\Hilvpi}{{\cal H}_{\vpi}}%
\newcommand{\Hilvpspri}{{\cal H}_{\vpspri}}%
\newcommand{\Hilvpshat}{{\cal H}_{\vpshat}}%
\newcommand{\Hilvpstil}{{\cal H}_{\vpstil}}%
\newcommand{\Hilvppm}{{\cal H}_{\vp \pm}}%
\newcommand{\Hilvpmp}{{\cal H}_{\vp \mp}}%
\newcommand{\Hilvpp}{{\cal H}_{\vp +}}%
\newcommand{\Hilvpm}{{\cal H}_{\vp -}}%
\newcommand{\Hilvpfp}{{\cal H}_{\vpf +}}%
\newcommand{\Hilvpfm}{{\cal H}_{\vps -}}%
\newcommand{\Hilvpfpm}{{\cal H}_{\vpf \pm}}%
\newcommand{\Hilvpfmp}{{\cal H}_{\vpf \mp}}%
\newcommand{\Hilvpspd}{{\hat{\cal H}}_{\vps +}}%
\newcommand{\Hilvpsp}{{\cal H}_{{\vpsp}}}%
\newcommand{\Hilvpsm}{{\cal H}_{{\vpsm}}}%
\newcommand{\Hilvpspm}{{\cal H}_{\vps \pm}}%
\newcommand{\Hilvpsmp}{{\cal H}_{\vps \mp}}%
\newcommand{\otimesHili}{\otimes_{i\ge 1}\Hili}%
\newcommand{\otimesOmei}{\otimes_{i\ge 1}\Omei}%
\newcommand{\otimesui}{\otimes_{i\ge 1}\ui}
\newcommand{\vpiset}{\{\vpi\}_{i=0,1,\cdots}}
\newcommand{\Hiltau}{{\cal H}_{\tau}}%
\newcommand{\HiltauN}{{\cal H}_{\tau}^{\N}}%
\newcommand{\GNSvpvec}{\mit\Omega_\vp}%
\newcommand{\omegaJ}{\omega_{\J}}%
\newcommand{\omeI}{\omega_{\I}}%
\newcommand{\omeJ}{\omega_{\J}}%
\newcommand{\Ome}{\mit\Omega}%
\newcommand{\Omef}{\Ome_{1}}%
\newcommand{\Omes}{\Ome_{2}}%
\newcommand{\Omei}{\Ome_{i}}%
\newcommand{\Omeo}{\Ome_{0}}%
\newcommand{\Omevp}{\Ome_{\vp}}%
\newcommand{\Omepsi}{\Ome_{\psi}}%
\renewcommand{\Vec}{\Ome}%
\newcommand{\Vecvp}{\Ome_{\vp}}%
\newcommand{\Vecvpf}{\Ome_{\vpf}}%
\newcommand{\Vecvps}{\Ome_{\vps}}%
\newcommand{\Vecvpsp}{\Ome_{\vpsp}}
\newcommand{\Vecvpspri}{\Ome_{\vpspri}}%
\newcommand{\Vecvpstil}{\Ome_{\vpstil}}%
\newcommand{\Vecd}{\widehat{\Ome}}%
\newcommand{\Vecds}{\widehat{\Ome}_{2}}%
\newcommand{\pisd}{\widehat{\pi}_{2}}%
\newcommand{\GNS}{\bigl\{\Hilvp, \pivp, \Vecvp \bigr\}}%
\newcommand{\proofs}{{\it{Proof.}}}%
\newcommand{\proofend}{{\hfill $\square$}\  \\}
\newcommand{\phivec}{{\mit \Phi}}%
\newcommand{\psivec}{{\mit \Psi}}%
\newcommand{\Afp}{A_{1+} }%
\newcommand{\Afm}{A_{1-} }%
\newcommand{\Afpm}{A_{1\pm} }%
\newcommand{\Asp}{A_{2+} }%
\newcommand{\Asppri}{A_{2+}^{\prime} }%
\newcommand{\Asm}{A_{2-} }%
\newcommand{\Aspm}{A_{2\pm} }%
\newcommand{\Aip}{A_{i +} }%
\newcommand{\Aim}{A_{i -} }%
\newcommand{\Aipm}{A_{i \pm} }%
\newcommand{\Afsig}{A_{1 \sig} }%
\newcommand{\Afsigp}{A_{1 {\sig^{\prime}}  } }%
\newcommand{\Assig}{A_{2 \sig} }%
\newcommand{\Assigp}{A_{2 {\sig^{\prime}}  } }%
\newcommand{\Alsig}{\Al_{ \sig} }%
\newcommand{\Alsigp}{\Al_{ {\sig^{\prime}}  } }%
\newcommand{\AlIsig}{\AlI_{ \sig} }%
\newcommand{\AlIsigp}{\AlI_{ {\sig^{\prime}}  } }%
\newcommand{\AlJsig}{\AlJ_{ \sig} }%
\newcommand{\AlJsigj}{\AlJ_{ \sig(j)} }%
\newcommand{\AlJsigp}{\AlJ_{ {\sig^{\prime}}  } }%
\newcommand{\AlIfsig}{\AlIf_{\sig} }%
\newcommand{\AlIfsigp}{\AlIf_{ {\sig^{\prime}}  } }%
\newcommand{\AlIssig}{\Als_{ \sig} }%
\newcommand{\AlIssigp}{\AlIs_{ {\sig^{\prime}}  } }%
%
%
\newcommand{\Alssigp}{\Als_{{\sig^{\prime}}  } }%
\newcommand{\ovp}{\overline{\vp}}%
\newcommand{\ovpf}{\overline{\vpf}}%
\newcommand{\ovps}{\overline{\vps}}%
\newcommand{\pivpAl}{\pivp(\Al)}%
\newcommand{\pivpI}{\pivp(\AlI)}%
\newcommand{\pivpIf}{\pivp\bigl(\AlIf\bigr)}%
\newcommand{\pivpIs}{\pivp\bigl(\AlIs\bigr)}%
\newcommand{\Svpfs}{S(\vpf,\,\vps)}%
\newcommand{\po}{p_{0}}%
\newcommand{\monomials}{\Af\cdots \Ak}
\newcommand{\piiAlIidoub}{\pii\bigl(\AlIi \bigr)^{\prime\prime}}
\newcommand{\BH}{\Bl(\Hil)}
\newcommand{\BHi}{\Bl(\Hili)}
\newcommand{\BHj}{\Bl(\Hilj)}
\newcommand{\BHo}{\Bl(\Hilo)}
\newcommand{\BHf}{\Bl(\Hilf)}
\newcommand{\BHs}{\Bl(\Hils)}
\newcommand{\BHvp}{\Bl(\Hilvp)}
\newcommand{\BHvpp}{\Bl(\Hilvpp)}
\newcommand{\BHvpm}{\Bl(\Hilvpm)}
\newcommand{\BHvppm}{\Bl(\Hilvppm)}
\newcommand{\BHvpf}{\Bl(\Hilvpf)}
\newcommand{\BHvps}{\Bl(\Hilvps)}
\newcommand{\BHvpsp}{\Bl(\Hilvpsp)}
%
\newcommand{\pivpAlI}{\pivp(\AlI)}
\newcommand{\pivpAlJ}{\pivp(\AlJ)}
\newcommand{\pivpAlIf}{\pivp(\AlIf)}
\newcommand{\pivpAlIs}{\pivp(\AlIs)}
\newcommand{\pivpAlIfp}{\pivp(\AlIfp)}
\newcommand{\pivpAlIfm}{\pivp(\AlIfm)}
\newcommand{\pivpAlIsp}{\pivp(\AlIsp)}
\newcommand{\pivpAlIsm}{\pivp(\AlIsm)}
\newcommand{\pivpAlIfus}{\pivp(\AlIfus)}
\newcommand{\pivpAlIfpm}{\pivp(\AlIfpm)}
\newcommand{\pivpAlIspm}{\pivp(\AlIspm)}
\newcommand{\pivpfAlIf}{\pivpf(\AlIf)}
\newcommand{\pivpfAlIfp}{\pivpf(\AlIfp)}
\newcommand{\pivpfAlIfm}{\pivpf(\AlIfm)}
\newcommand{\pivpfAlIfpm}{\pivpf(\AlIfpm)}
\newcommand{\wpri}{\prime\prime}%
\newcommand{\pivpAlIwpri}{\pivp(\AlI)^{\wpri}}
\newcommand{\pivpAlIfwpri}{\pivp(\AlIf)^{\wpri}}
\newcommand{\pivpAlIswpri}{\pivp(\AlIs)^{\wpri}}
\newcommand{\pivpAlIfswpri}{\pivp(\AlIfs)^{\wpri}}
%
\newcommand{\pivpAlIfpwpri}{\pivp(\AlIfp)^{\wpri}}
\newcommand{\pivpAlIfmwpri}{\pivp(\AlIfm)^{\wpri}}
\newcommand{\pivpAlIspwpri}{\pivp(\AlIsp)^{\wpri}}
\newcommand{\pivpAlIsmwpri}{\pivp(\AlIsm)^{\wpri}}
\newcommand{\pivpAlIfpmwpri}{\pivp(\AlIfpm)^{\wpri}}
\newcommand{\pivpAlIspmwpri}{\pivp(\AlIspm)^{\wpri}}
\newcommand{\pivpfAlIfwpri}{\pivpf(\AlIf)^{\wpri}}
\newcommand{\pivpfAlIfpwpri}{\pivpf(\AlIfp)^{\wpri}}
\newcommand{\pivpfAlIfmwpri}{\pivpf(\AlIfm)^{\wpri}}
\newcommand{\pivpfAlIfpmwpri}{\pivpf(\AlIfpm)^{\wpri}}
%
\newcommand{\pivpsAlIs}{\pivps(\AlIs)}
\newcommand{\pivpsAlIsp}{\pivps(\AlIsp)}
\newcommand{\pivpsAlIswpri}{\pivps(\AlIs)^{\wpri}}
\newcommand{\pivpsAlIspwpri}{\pivps(\AlIsp)^{\wpri}}
\newcommand{\pivpsAlIsmwpri}{\pivps(\AlIsm)^{\wpri}}
\newcommand{\pivpsAlIspmwpri}{\pivps(\AlIspm)^{\wpri}}
%
\newcommand{\pivpspAlIswpri}{\pivpsp(\AlIs)^{\wpri}}
\newcommand{\pivpspAlIspwpri}{\pivpsp(\AlIsp)^{\wpri}}
\newcommand{\pivpspAlIsmwpri}{\pivpsp(\AlIsm)^{\wpri}}
\newcommand{\pivpspAlIspmwpri}{\pivpsp(\AlIspm)^{\wpri}}
\newcommand{\pivpspAlIspri}{\pivpsp(\AlIs)^{\pri}}
\newcommand{\pivpspAlIsppri}{\pivpsp(\AlIsp)^{\pri}}
\newcommand{\pivpspAlIsmpri}{\pivpsp(\AlIsm)^{\pri}}
\newcommand{\pivpspAlIspmpri}{\pivpsp(\AlIspm)^{\pri}}
\newcommand{\pistilAlIswpri}{\pistil(\AlIs)^{\wpri}}
\newcommand{\pistilAlIspwpri}{\pistil(\AlIsp)^{\wpri}}
\newcommand{\pistilAlIsmwpri}{\pistil(\AlIsm)^{\wpri}}
\newcommand{\pistilAlIspmwpri}{\pistil(\AlIspm)^{\wpri}}
\newcommand{\pri}{\prime}%
\newcommand{\pivpAlIpri}{\pivp(\AlI)^{\pri}}
\newcommand{\pivpAlIfpri}{\pivp(\AlIf)^{\pri}}
\newcommand{\pivpAlIspri}{\pivp(\AlIs)^{\pri}}
\newcommand{\pivpAlIfuspri}{\pivp(\AlIfus)^{\pri}}
\newcommand{\pisAlIs}{\pis(\AlIs)}
\newcommand{\pisAlIspri}{\pis(\AlIs)^{\pri}}
\newcommand{\pisAlIswpri}{\pis(\AlIs)^{\wpri}}
\newcommand{\pisdAlIswpri}{\pisd(\AlIs)^{\wpri}}
\newcommand{\pisdAlIspri}{\pisd(\AlIs)^{\pri}}
\newcommand{\pivpAlIfppri}{\pivp(\AlIfp)^{\pri}}
\newcommand{\pivpAlIfmpri}{\pivp(\AlIfm)^{\pri}}
\newcommand{\pivpAlIsppri}{\pivp(\AlIsp)^{\pri}}
\newcommand{\pivpAlIsmpri}{\pivp(\AlIsm)^{\pri}}
\newcommand{\pivpAlIfpmpri}{\pivp(\AlIfpm)^{\pri}}
\newcommand{\pivpAlIspmpri}{\pivp(\AlIspm)^{\pri}}
\newcommand{\pivpfAlIfpri}{\pivpf(\AlIf)^{\pri}}
\newcommand{\pivpfAlIfppri}{\pivpf(\AlIfp)^{\pri}}
\newcommand{\pivpfAlIfmpri}{\pivpf(\AlIfm)^{\pri}}
\newcommand{\pivpfAlIfpmpri}{\pivpf(\AlIfpm)^{\pri}}
\newcommand{\pivpA}{\pivp(A)}
\newcommand{\pivpAf}{\pivp(\Af)}
\newcommand{\pivpAs}{\pivp(\As)}
\newcommand{\pivpAlfp}{\pivp(\Afp)}
\newcommand{\pivpAfm}{\pivp(\Afm)}
\newcommand{\pivpAsp}{\pivp(\Asp)}
\newcommand{\pivpAsm}{\pivp(\Asm)}
\newcommand{\pivpAfpm}{\pivp(\Afpm)}
\newcommand{\pivpAspm}{\pivp(\Aspm)}
\newcommand{\pivpfAf}{\pivpf(\Af)}
\newcommand{\pivpfAfp}{\pivpf(\Afp)}
\newcommand{\pivpfAfm}{\pivpf(\Afm)}
\newcommand{\pivpfAfpm}{\pivpf(\Afpm)}
\newcommand{\pivpAt}{\pivp(\Theta(A))}
\newcommand{\pivpAft}{\pivp(\Theta(\Af))}
\newcommand{\pivpAst}{\pivp(\Theta(\As))}
\newcommand{\pivpAfpt}{\pivp(\Theta(\Afp))}
\newcommand{\pivpAfmt}{\pivp(\Theta(\Afm))}
\newcommand{\pivpAspt}{\pivp(\Theta(\Asp))}
\newcommand{\pivpAsmt}{\pivp(\Theta(\Asm))}
\newcommand{\pivpAfpmt}{\pivp(\Theta(\Afpm))}
\newcommand{\pivpAspmt}{\pivp(\Theta(\Aspm))}
\newcommand{\pivpfAft}{\pivpf(\Theta(\Af))}
\newcommand{\pivpfAfpt}{\pivpf(\Theta(\Afp))}
\newcommand{\pivpfAfmt}{\pivpf(\Theta(\Afm))}
\newcommand{\pivpfAfpmt}{\pivpf(\Theta(\Afpm))}
\newcommand{\Vs}{V_{2}}
\newcommand{\Tr}{\mathbf{Tr}}%
%
\begin{abstract}
The problem of existence and uniqueness of a state of 
 a joint system with given restrictions to subsystems
 is studied for a Fermion system, where a novel feature 
 is non-commutativity between algebras of subsystems.
 
For an arbitrary (finite or infinite) number of  given 
 subsystems, a product state extension is shown to exist if and only if 
all states of subsystems except at most one are  even
(with respect to the Fermion number).
If the states of all subsystems are pure, then the same
 condition is shown to be necessary and sufficient for the existence
 of  any joint extension.
If the condition holds,  the unique product 
state  extension is the only joint extension. 

For a pair of subsystems, with one of the given subsystem
 states pure, a necessary and sufficient condition for the existence of a
 joint extension and the form of all joint extensions
 (unique for almost all cases) are given.
 For a pair of subsystems with non-pure subsystem states,
 some  classes  of examples of joint extensions are 
 given where non-uniqueness of joint extensions prevails.
\end{abstract}
%
%
%
%
\section{Introduction and Results}
\label{sec:INTRO}
The problem of extending given pure states of a pair of subsystems
to a state of the joint system has been treated 
 for a bipartite CAR system 
by one of the authors \cite{MORIYAentangle} 
in connection with the subject of entanglement in quantum information theory
 where one studies mutual relations of states 
of subsystems obtained as restrictions
 of a pure state of the joint system.

In the present article, we continue the study of 
 a joint extension of states of subsystems to a state of the 
joint system for a Fermion system.

We consider a $\cstar$-algebra $\Al$, called a CAR algebra
 or a Fermion algebra, which is generated by its elements 
$\ai$ and $\aicr$, $i\in \NN$ ($\NN=\{1,2,\cdots\}$) satisfying the following
 canonical anticommutation relations(CAR).
\begin{eqnarray}
\label{eq:CAR1}
\{ \aicr, \aj \}&=&\delta_{i,j}\, \id  \\
\label{eq:CAR2}
\{ \aicr, \ajcr \}&=&\{ \ai, \aj \}=0,
\end{eqnarray}
($i,j\in \NN$), 
where $\{A, B\}=AB+BA$ (anticommutator) and 
 $\delta_{i,j}=1$ for  $i = j$ and 
 $\delta_{i,j}=0$ otherwise.
For any  subset $\I$ of $\NN$, $\AlI$ denotes
the $\cstar$-subalgebra
 generated by 
$\ai$ and $\aicr$, $i \in \I$. 

As subsystems, we consider $\AlI$ with mutually disjoint 
subsets $\I$'s.
For a pair of disjoint subsets $\If$ and $\Is$ of $\NN$, let 
$\vpf$ and $\vps$ be given states of $\AlIf$ and $\AlIs$,
 respectively.
 If a state $\vp$ of the joint system $\AlIfs$
 (which is the same as the $\cstar$-subalgebra of $\Al$
 generated by $\AlIf$ and $\AlIs$)
 coincides with $\vpf$ on $\AlIf$ and $\vps$ on $\AlIs$,
 i.e., 
\begin{eqnarray*}
\vp(\Af)&=&\vpf(\Af),\quad \Af \in \AlIf,\nonum \\
\vp(\As)&=&\vps(\As),\quad \As \in \AlIs,
\end{eqnarray*}
then $\vp$ is called a joint extension of $\vpf$ and $\vps$.
 As a special case, 
 if 
\begin{eqnarray}
\label{eq:EXTpair}
\vp(\Af \As)=\vpf(\Af) \vps(\As)
\end{eqnarray}
 holds for all $\Af \in \AlIf$ and 
 all $\As  \in \AlIs$, then 
$\vp$ is called a product state extension of $\vpf$ and $\vps$.
For an arbitrary (finite or infinite)
 number of subsystems, 
 $\AlIf$, $\AlIs,\cdots$ with mutually disjoint $\I$'s 
 and  a set of given states $\vpi$ of $\AlIi$,
 a state $\vp$ of $\AlcupIi$ is called a product 
 state extension if it satisfies
\begin{eqnarray}
\label{eq:EXTk}
\vp(\Af\As \cdots \Ak)=\prod_{i=1}^{k}\vpi(\Ai), \quad \Ai \in \AlIi,
\end{eqnarray}
for all $k$.

A crucial role
 is played by the unique 
 automorphism $\Theta$
 of $\Al$ 
characterized by 
\begin{eqnarray}
\label{eq:THETA}
\Theta(\ai)=-\ai,\quad  \Theta(\aicr)=-\aicr 
\end{eqnarray}
for all $i \in \NN$.
 The even and odd parts of $\Al$ and $\AlI$ are defined by
\begin{eqnarray}
\label{eq:AlpmEQ}
\Alpm &\equiv& \bigl\{ A \in \Al \, |\, \Theta(A)=\pm A  \bigr\}, \\
\AlIpm &\equiv& \Alpm\cap \AlI. 
\end{eqnarray}

For any $A \in \Al$ (or $\AlI$), we have the following decomposition
\begin{eqnarray}
\label{eq:ApmEQ}
A=\Ap+\Am,\quad \Apm=\frac{1}{2}\bigl(A\pm \Theta(A)\bigr)\in \Alpm\,\,
\text{(or}\,\, \AlIpm \text{)}.
\end{eqnarray}

A state $\vp$ of $\Al$ or $\AlI$ is called even
 if it is $\Theta$-invariant:
\begin{eqnarray}
\label{eq:evenstateEQ}
\vp\bigl( \Theta(A)\bigr)=\vp(A)
\end{eqnarray}
 for all $A \in \Al$ (or $A \in \AlI$).
Note that 
$\vp(A)=0$ for all $A \in \Alm\  (\AlIm$) is equivalent 
to the condition that $\vp$ is an even state of $\Al\  (\AlI)$. 
%

For a state of a $\cstar$-algebra $\Al$ ($\AlI$),
 $\GNS$ denotes the GNS triplet of a Hilbert space 
$\Hilvp$, a representation $\pivp$ of $\Al$
 (of $\AlI$), and a vector $\Vecvp\in \Hilvp$, which is cyclic for
 $\pivp(\Al)$ ($\pivp(\AlI)$)
 and satisfies 
\begin{eqnarray*}
\vp(A)&=&(\Vecvp,\, \pivp(A)\Vecvp)
\end{eqnarray*}
 for all $A \in \Al$ ($\AlI$).  For any 
$x\in \Bl(\Hilvp)$, we write 
\begin{eqnarray*}
{\overline{\vp}}(x)=(\Vecvp,\, x\Vecvp).
\end{eqnarray*}

The first group of our results are the following 
 three theorems related to a 
 product state extension.
\begin{thm}
\label{thm:PRODUCT}
Let $\If, \Is, \cdots$ be an arbitrary (finite or infinite)
number of mutually disjoint subsets of $\NN$
 and $\vpi$ be a given state of $\AlIi$ for each $i$. \\
{\rm{(1)}} A product state extension of $\vpi$, $i=1,2,\cdots,$
 exists if and only if all states $\vpi$
 except at most one are   even.
It is unique if it exists. It is even if and only if all $\vpi$
 are even. \\
{\rm{(2)}}  Suppose that  all  $\vpi$ are pure. 
If there  exists a  joint extension of $\vpi$, $i=1,2,\cdots,$
then  all states $\vpi$ except at most one have to be  even.
 If this is the case, the joint extension is uniquely given 
 by the product state extension and is a pure state.
\end{thm}
\begin{rem}
 In Theorem \ref{thm:PRODUCT} (2), the product state property
 (\ref{eq:EXTk}) is not assumed but it is derived from the 
 purity assumption for all $\vpi$.
\end{rem}

The purity of all $\vpi$ does not follow from that  of their joint 
 extension $\vp$ in general.
 For a product state extension $\vp$, however, we have the 
 following two theorems about consequences of purity 
 of $\vp$.
%

\begin{thm}
\label{thm:PROEXT-1}
Let $\vp$ be the product state extension of states $\vpi$
 with disjoint $\Ii$.
Assume that all $\vpi$ except $\vp_1$  are even. \\
{\rm{(1)}}   $\vpf$ is pure 
 if   $\vp$ is  pure.\\
{\rm{(2)}} Assume that  $\pivpf$ and $\pivpft$ are not disjoint.
Then $\vp$ is pure if and only if 
all $\vpi$ are pure.
 In particular, this is the case  if $\vp$ is even. 
\end{thm}
%
\begin{rem}
If $\If$ is finite,
  the assumption of  Theorem \ref{thm:PROEXT-1}
 (2) holds and hence the  conclusion follows automatically.
\end{rem}

In the case not covered by Theorem \ref{thm:PROEXT-1},
 the following result gives a complete analysis
 if we take
 $\cup_{i\ge 2}\Ii$ in Theorem \ref{thm:PROEXT-1}
 as one subset of $\NN$.
\begin{thm}
\label{thm:PROEXT-D}
Let $\vp$ be the product state extension of states $\vpf$
 and $\vps$ of $\AlIf$ and $\AlIs$ 
 with disjoint $\If$ and $\Is$
 where $\vps$ is even and $\vpf$ is such that 
$\pivpf$ and $\pivpft$ are disjoint.\\
{\rm{(1)}} $\vp$ is pure if and only if $\vpf$ and the restriction 
$\vpsp$ of $\vps$ to $\AlIsp$
 are both pure.\\
{\rm{(2)}} Assume that $\vp$ is pure.
 $\vps$ is not pure if and only if 
\begin{eqnarray}
 \label{eq:PROEXT-D3}
 \vps=\frac{1}{2}(\hvps+\hvps\Theta)
 \end{eqnarray}
 where $\hvps$ is pure and $\pihvps$ and $\pihvpst$
 are disjoint.
\end{thm}
\begin{rem}
 The first  two theorems are some generalization of results 
 in \cite{POWERS}
  with the following overlap.
The first part of Theorem \ref{thm:PRODUCT} (1) 
is given in \cite{POWERS} as Theorem 5.4
(the if part and  uniqueness) and a discussion after Definition  5.1
 (the only if part).
Theorem \ref{thm:PRODUCT} (2) and Theorem 
 \ref{thm:PROEXT-1} are 
given in Theorem 5.5 of \cite{POWERS}
 under the assumption that  all $\vpi$
 are even.
 Since the reference \cite{POWERS} does not seem to be widely available,
 we present a complete proof in $\S$ \ref{sec:PSE}.
The if part of the first part of 
Theorem \ref{thm:PRODUCT} (1) is also given in Theorem 11.2 of 
 \cite{ARAKIMORIYA} which plays a crucial role in that paper.
\end{rem}

The rest of our results concerns a joint extension of states
 of two subsystems,
 not satisfying 
 the product state property (\ref{eq:EXTpair}).
We  need a few  more notation. 
For two states $\vp$ and $\psi$ of a $\cstar$-algebra $\AlIf$,
 consider any representation $\pi$ of $\AlIf$ on a Hilbert space
 $\Hil$ containing  vectors $\phivec$ and  $\psivec$
 such that 
\begin{eqnarray}
\label{eq:phivecEQ}
\vp(A)=(\phivec,\, \pi(A) \phivec),\quad
\psi(A)=(\psivec,\, \pi(A)\psivec).
\end{eqnarray}
The transition probability between 
 $\vp$ and $\psi$ is defined (\cite{UHLMANN77}) by
\begin{eqnarray}
\label{eq:Trans0}
P(\vp,\,\psi)\equiv
\sup | ({\phivec},\,\psivec) |^{2}
\end{eqnarray}
where the supremum is taken over all $\Hil$, $\pi$, 
 $\phivec$ and $\psivec$ as described 
 above.
 For a state $\vpf$ of $\AlIf$, we need the following quantity
\begin{eqnarray*}
p(\vpf)\equiv \Pvpfvpft^{1/2}
\end{eqnarray*}
where 
$\vpf\Theta$ denotes the state 
$\vpf\Theta(A)=\vpf(\Theta(A))$, $A\in \AlIf$.

If $\vpf$ is pure, then $\vpf\Theta$ is also pure 
 and the representations $\pivpf$ and $\pivpft$
 are both irreducible. There are two alternatives.\\
\ \\
\quad $(\alpha)$ They are unitarily equivalent. \\
\quad $(\beta)$ They are mutually disjoint. In this case $p(\vpf)=0$.\\
\ \\
In  the  case ($\alpha$),  
 there exists a self-adjoint unitary $\uone$ on $\Hilvpf$
 such that 
\begin{eqnarray} 
\label{eq:uone1}
\uone \pivpf(A) \uone&=&\pivpf(\Theta(A)),\quad A\in \AlIf,\\
\label{eq:uone2}
(\Vecvpf,\,\uone\Vecvpf)&\ge&0,
\end{eqnarray}
by Lemma \ref{lem:uone} in $\S$ \ref{sec:PSE}.

For two states $\vp$ and $\psi$, we introduce 
\begin{eqnarray}
\label{eq:lamvppsiEQ}
\lamvppsi\equiv\sup\bigl\{\lam \in \R;\ \vp-\lam\psi\ge0\bigr\}
\end{eqnarray}
Since $\vp-\lam_{n}\psi\ge 0$
 and $\lim \lam_{n}=\lam$ imply $\vp-\lam \psi\ge 0$,
 we have 
\begin{eqnarray}
\label{eq:lamineEQ}
\vp \ge\lamvppsi \psi.
\end{eqnarray}
We need 
\begin{eqnarray}
\label{eq:lamvpsEQ}
\lamvps \equiv \lam(\vps,\,\vps \Theta).
\end{eqnarray}
The next Theorem 
provides a complete answer 
for a joint extension $\vp$
 of states $\vpf$ and $\vps$ of $\AlIf$ and $\AlIs$, 
 when one of them is pure.
 It will be shown  in $\S$ \ref{sec:PUREGENERAL}.
\begin{thm}
\label{thm:FPURE}
Let $\vpf$ and $\vps$ be states of $\AlIf$ and $\AlIs$
 for  disjoint subsets 
$\If$ and $\Is$.
Assume that $\vpf$ is pure.\\
{\rm{(1)}} A joint 
extension $\vp$ of $\vpf$ and  $\vps$
 exists 
  if and only if 
\begin{eqnarray}
\label{eq:EXTok}
\lamvps \ge \frac{1-\pvpf}{1+\pvpf}.
\end{eqnarray}
{\rm{(2)}} If (\ref{eq:EXTok}) holds and if $\pvpf\ne0$,
 then a  joint extension $\vp$ is unique and satisfies
\begin{eqnarray}
\label{eq:EXTform1}
\vp(\Af \As)&=&\vpf(\Af)\vps(\Asp)+\frac{1}{\pvpf}
f(\Af)\vps(\Asm),\\
\label{eq:EXTform2}
f(\Af)&\equiv& \ovpf(\pivpf(\Af)\uone)
\end{eqnarray}
for $\Af\in \AlIf$  and $\As=\Asp+\Asm$,
 $\Aspm \in \AlIspm$.\\
{\rm{(3)}} If  $\pvpf=0$, (\ref{eq:EXTok})
 is equivalent  to evenness of $\vps$.
If this is the case, at least a product state extension of 
Theorem \ref{thm:PRODUCT} exists. \\
{\rm{(4)}} Assume that  $\pvpf=0$ and 
 $\vps$ is even. There exists a joint extension 
 of $\vpf$ and $\vps$ other than the unique product state extension
 if and only if $\vpf$ and $\vps$ satisfy 
 the following pair of conditions$:$\\
\quad {\rm{(4-i)}} $\pivpf$ and $\pivpft$ are unitarily equivalent.\\
\quad {\rm{(4-ii)}} There exists a state $\vpstil$ 
 of $\AlIs$ such that $\vpstil \ne \vpstilt$
 and 
\begin{eqnarray}
\label{eq:2dec}
\vps=\frac{1}{2}\bigl(\vpstil +\vpstilt \bigr).
\end{eqnarray}
{\rm{(5)}} If  $\pvpf=0$, then corresponding to each $\vpstil$
 above, there exists a joint extension $\vp$
 which satisfies 
\begin{eqnarray}
\label{eq:vppriform}
\vp(\Af \As)=\vpf(\Af)\vps(\Asp)+
\vpfb(\pivpf(\Af)\uone)\vpstil(\Asm).  
\end{eqnarray}
Such extensions along with the 
 unique product state extension (which satisfies 
(\ref{eq:vppriform}) for 
 $\vpstil=\vps$) exhaust all  joint extensions 
 of $\vpf$ and $\vps$
 when $\pvpf=0$.
\end{thm}
\begin{rem}
The condition (\ref{eq:EXTok})
 is sufficient for the existence
 of a joint extension also for general 
 states $\vpf$ and $\vps$.
A continuation of this work including this result
is under preparation.
\end{rem}

We have a necessary and sufficient condition for the existence 
 of a joint extension 
 of  states $\vpf$ and $\vps$
 under a specific condition on $\vpf$.
  Proof will be given in $\S$ \ref{sec:F-NONPURE}.
\begin{thm}
\label{thm:F-NONPURE}%
Let $\vpf$ and $\vps$ be states of $\AlIf$ and $\AlIs$
 for  disjoint subsets 
$\If$ and $\Is$.  Assume that 
$\pivpf$ and $\pivpft$
 are disjoint. Then a joint extension of $\vpf$
 and $\vps$ 
 exists if and only if $\vps$ is even.
\end{thm}

Theorem \ref{thm:FPURE}
and Theorem \ref{thm:F-NONPURE}
 are not symmetric in 
 $\vpf$ and $\vps$. The following examples provide methods of construction 
  (of joint extensions)
  which are  symmetric in $\vpf$ and $\vps$. 

{\underline{\it{Example 1}}}\\ 
 Let $\If$ and $\Is$ be mutually 
 disjoint finite subsets of $\NN$.
 Let $\vrho\in \AlIfus$ be an invertible density 
 matrix, 
 namely $\vrho\ge \lam \id$ for some
 $\lam>0$ and $\Tr(\vrho)=1$, where $\Tr$
 denotes the matrix trace on $\AlIfus$.
 Take any $x=x^{\ast}\in \AlIfm$
 and  $y=y^{\ast}\in \AlIsm$
 satisfying 
$\Vert x \Vert \Vert y \Vert \le \lam$.
 Let 
$\vpf(\Af)\equiv\Tr(\vrho \Af)$ for $\Af \in \AlIf$
and $\vps(\As)\equiv \Tr(\vrho \As)$ 
  for $\As \in \AlIs$. 
Then 
  \begin{eqnarray}
\label{eq:vrhopriEQ}
\vp_\vrhopri(A)\equiv \Tr(\vrhopri A),\quad 
\vrhopri\equiv \vrho+i xy.
\end{eqnarray}
 for $A \in \AlIfus$
is a state of $\AlIfus$
 and has $\vpf$ and $\vps$ as its restrictions to 
$\AlIf$ and $\AlIs$, irrespective of the choice
 of $x$ and $y$ satisfying the above conditions.

{\underline{\it{Example 2}}}\\ 
 Let $\If$ and $\Is$ be mutually 
 disjoint  subsets of $\NN$.
 Let $\vp$ and $\psi$ be states of $\AlIf$ and 
 $\AlIs$ such that
  \begin{eqnarray*}
 \vp=\sum_{i}\lami \vpi,\quad
 \psi=\sum_{i}\lami \psii,\ 
(0< \lami,\ \sum_{i}\lami=1),
\end{eqnarray*}
 where $\vpi$ and 
 $\psii$ are states of $\AlIf$ and $\AlIs$ 
 which have a joint extension $\chii$ for each 
$i$.
\begin{eqnarray*}
\chi=\sum_{i} \lami \chii
\end{eqnarray*}
 is a joint extension of $\vp$ and $\psi$.

This simple example yields next more elaborate  ones.

{\underline{\it{Example 3}}}\\
 Let $\vp$ and  $\psi$ be states of $\AlIf$
 and $\AlIs$ 
 for disjoint $\If$ and $\Is$
with  (non-trivial) decompositions
 \begin{eqnarray*}
 \vp=\lam\vpf+(1-\lam)\vps,\quad
 \psi=\mu\psif+(1-\mu)\psis,\quad
(0< \lam,\mu<1)
\end{eqnarray*}
 where $\vpf$ and $\vps$ are even.
Product state extensions $\vpi\psij$
 of $\vpi$ and $\psij$ yield
\begin{eqnarray}
 \chi&\equiv &(\lam \mu+\kappa)\vpf\psif
+(\lam(1-\mu) -\kappa)\vpf\psis \nonum\\
&&((1-\lam)\mu-\kappa)\vps\psif
+((1-\lam)(1-\mu) +\kappa)\vps\psis,
\end{eqnarray}
which  is a joint extension of $\vp$
  and $\psi$ for all $\kappa\in \R$
 satisfying 
\begin{eqnarray}
-\min (\lam \mu,\,(1-\lam)(1-\mu))\le \kappa \le 
 \min ((1-\lam) \mu,\,\lam(1-\mu)).
\end{eqnarray}

{\underline{\it{Example 4}}}\\ 
 Let $\vpk$, $k=1,\cdots,m$ and 
$\psil$, $l=1,\cdots,n$ be 
 states of $\AlIf$ and $\AlIs$
 for disjoint $\If$ and $\Is$.
Let 
\begin{eqnarray*}
\vp=\sum_{k=1}^{m}\lamk \vpk,\quad 
\psi=\sum_{l=1}^{n}\mul \psil
\end{eqnarray*}
 with $\lamk,\,\mul > 0$, $\sum \lamk=\sum \mul=1$.
 Assume that there exists 
 a joint extension $\chikl$ of $\vpk$
 and $\psil$ for each $k$ and $l$.
Then 
\begin{eqnarray}
\label{eq:extchi}
\chi=\sum_{k l} (\lamk \mul+\kappa_{kl})\chikl
\end{eqnarray}
 is a joint extension if 
\begin{eqnarray}
 (\lamk \mul+\kappa_{kl})\ge 0,\quad \sum_{l}\kappa_{kl}
=\sum_{k}\kappa_{kl}=0.
\end{eqnarray}
 Since the constraint for $mn$ parameters $\{\kappa_{kl}\}$
 are effectively $m+n-1$ linear relations
(because $\sum_{kl}\kappa_{kl}=0$ is common for 
$\sum_{l}\kappa_{kl}=0$ and $\sum_{k}\kappa_{kl}=0$ ),
 we have $mn-(m+n-1)=(m-1)(n-1)$ parameters for the joint extension 
(\ref{eq:extchi}).
%
\section{The Fermion Algebra}
\label{sec:NOTATION}
As already explained, we investigate the $\cstar$-algebra 
$\Al$ generated by $\ai$ and $\aicr$, $i \in \NN$,
 satisfying CAR (\ref{eq:CAR1}) and (\ref{eq:CAR2}), 
and its subalgebras $\AlI$, $\I\subset \NN$ generated 
 by $\ai$ and $\aicr$, $i \in \I$.

An important role is played by 
 the splitting of $\AlI$
 (for each $\I$) into $\Theta$-even and $\Theta$-odd parts by the 
formula (\ref{eq:ApmEQ}):
\begin{eqnarray}
\label{eq:AlIpmEQ}
\AlI=\AlIp+\AlIm.
\end{eqnarray}
\begin{rem}
 The notation $\AlIp$
 for the $\Theta$-even part of $\AlI$ should not be confused 
 with the set of all positive  elements of $\AlI$.
\end{rem}
\begin{lem}
\label{lem:epsCOMMUT}
 If $\If$ and $\Is$ are disjoint subsets of $\NN$
 and if $\Afpm\in \AlIfpm$
 and $\Aspm \in \AlIspm$,
 then
\begin{eqnarray}
\label{eq:epsCOMMUT}
\Afsig \Assigp=\vareps(\sigma,\,\sigma^{\prime})
\Assigp 
\Afsig
\end{eqnarray}
for $\sigma=\pm$ and $\sigma^{\prime}=\pm$
 where 
\begin{eqnarray}
\vareps(\sigma,\,\sigma^{\prime}) &=&-1,\  {\text{if}}\ 
\sigma=\sigma^{\prime}=-,\nonum \\
\vareps(\sigma,\,\sigma^{\prime}) &=&+1,\  {\text{otherwise}}.  
\end{eqnarray}
\end{lem}
\proofs\ 
Even degree monomials of $\ai$ and $\aicr$, $i \in \I$,
 are in $\AlIp$ and their linear span is dense in $\AlIp$.
Odd ones 
 are in $\AlIm$ and their linear span is dense in $\AlIm$.
 Therefore (\ref{eq:epsCOMMUT}) follows from (\ref{eq:CAR2}).
\proofend
\begin{rem}
We may rephrase (\ref{eq:epsCOMMUT}) by saying that 
$\AlIfm$ and
$\AlIsm$ anticommute while other pairs
 of $\AlIsig$ and $\AlIssigp$ commute.
\end{rem}

The algebra $\Ali$ for a one point subset $\I=\{i\}$,
 $i \in \NN$ is a linear span of the following 
four self-adjoint unitaries and is isomorphic to 
the algebra of all  $2\times 2$ matrices.
\begin{eqnarray}
\label{eq:unit0}
\uiz\equiv\id, \uih\equiv\ai+\aicr, \uif\equiv i(\ai-\aicr),
 \uim\equiv \aicr\ai-\ai\aicr.
\end{eqnarray}
In fact (\ref{eq:CAR1}) and (\ref{eq:CAR2}) 
imply  the multiplication
 rule of Pauli spin matrices among them. In addition,
\begin{eqnarray}
\label{eq:unit1}
\uiz, \uim \in \Alip,\quad  
\uih, \uif \in \Alim.
\end{eqnarray}
Hence monomials of $\uiji$ with distinct indices 
$i\in \I$ have a linear span dense in $\AlI$,
 monomials of even and odd total degrees 
in  $\uiji$, $\ji=1,2$, 
having a linear span dense in $\AlIpm$, respectively.
\begin{lem}
\label{lem:tech2}
For disjoint $\If$ and $\Is$, 
let  $\vp$ be a state of $\AlIfus$
 with its restrictions 
$\vpf$ and $\vps$  to $\AlIf$
 and $\AlIs$.
Then the representation $\pivp$ of $\AlIf$
 is quasi-equivalent to $\pivpf\oplus \pivpft$. 
\end{lem}
\proofs
\ 
 Let 
\begin{eqnarray}
\label{eq:HilvppmEQ}
\Hilvppm \equiv {\overline{\pivp(\AlIf)\pivp(\AlIspm)}}\Vecvp .
\end{eqnarray}
They are $\pivp(\AlIf)$ invariant.
 Let 
\begin{eqnarray*}
\Uji \equiv \prod_{i\in \Is} \uiji
\end{eqnarray*}
 where $\ji\ne0$ only for a finite number of indices  $i$
 and the order of the product is in the increasing  order of $i$
 from left to right.
 Let $\sigji=\pm$ according to whether the number of 
 $i\in \I_2$ with $\ji\in \{1,2\}$
 is even or odd.
 Then 
\begin{eqnarray*}
\Uji \in \AlIs_{\sigji}.
\end{eqnarray*}
The linear span of all $\Uji$ is dense in  $\AlIs$
 and the linear spans of 
$\pivp\bigl(\AlI \Uji\bigr) \Vecvp$ 
 with $\sigji=\pm$ are dense in $\Hilvppm$, respectively.

Due to (\ref{eq:epsCOMMUT}), we have
\begin{eqnarray*}
\bigl(\pivp(\Uji)\Vecvp,\, \pivp(\Af)\pivp(\Uji)\Vecvp\bigr)
&=&\vpf(\Af),\ {\text{if}}\ \sigji=+, \\
\bigl(\pivp(\Uji)\Vecvp,\, \pivp(\Af)\pivp(\Uji)\Vecvp\bigr)
&=&\vpf(\Theta(\Af)),\ \text{if}\ \sigji=-
\end{eqnarray*}
 for $\Af \in \AlIf$.
 Therefore $\pivp(\Af)$, $\Af \in \AlIf$, 
 restricted to $\Hilvpp$ and to $\Hilvpm$
 are quasi-equivalent to $\pivpf$ and $\pivpft$, respectively.
Since $\Hilvpp+\Hilvpm$ is dense in $\Hilvp$,
 $\pivp$ is quasi-equivalent to $\pivpf\oplus \pivpft$.
\proofend 
\begin{cor}
\label{cor:perp}
If $\pivpf$ and $\pivpft$ are disjoint, then
\begin{eqnarray}
\label{eq:perpEQ}
\Hilvpp\perp \Hilvpm,
\end{eqnarray}
 and   
  $\pivp$ of $\Al(\I_{1})$ restricted to 
 $\Hilvppm$ are quasi-equivalent to 
 $\pivpf$ and $\pivpft$.
\end{cor}

We will use the following Lemma repeatedly.
 It is Lemma 4.11 in Chapter IV of \cite{TAKESAKI1}
 and Lemma 2-2 in \cite{POWERScarges}.
 According to the latter, it is due to Guichardet.
\begin{lem}
\label{lem:TAKESAKI4.11}
If $\Bl$ is a $\cstar$-subalgebra of a $\cstar$-algebra  $\Cl$
 and if the restriction $\omeB$ of a state $\ome$ of $\Cl$
 to $\Bl$ is a pure state, then
\begin{eqnarray}
\label{eq:TAKESAKI4.11}
\ome(xy)=\ome(x)\ome(y),\quad  x\in \Bl, y \in \Bl^{\prime}\cap \Cl.
\end{eqnarray}
\end{lem}
\section{Product State Extension}
\label{sec:PSE}
\subsection{Theorem \ref{thm:PRODUCT} (1)}
\label{subsec:PSE1}
 
(a) {\underline{A concrete construction}}\\
We will give a (concrete) representation and 
a representative vector for  the state $\vp$ which is 
 a product state extension of $\vpi$, $i=1,2,\cdots$.

Since the numbering of subsystems are irrelevant,
 we assume that the indices are $0,1,2,\cdots$ and 
 $\vpf$, $\vps$, $\cdots$ are all even,
 while $\vpo$ need not be even, 
 where $\vpi$ is a state of $\AlIi$, $i\ge 0$.
Let $\I=\cup_{i\ge 0}\Ii$.

Let $(\Hili$, $\pii$, $\Omei)$ be the GNS triplet for $\vpi$,
 $i=0,1,2,\cdots$.
 Since $\vpi$ is assumed to be even for $i \ge 1$,
 there exists a unitary oprtator $\ui$ on $\Hili$ such that 
\begin{eqnarray}
\label{eq:ui1}
\ui \pii(\Ai)\Omei = \pii \bigl(\Theta(\Ai)\bigr)\Omei,\quad \Ai\in \AlIi.
\end{eqnarray}
(This defines an isomorphic operator with the dense 
 domain  $\pii (\AlIi) \Omei$ and the same 
 range and its closure defines a unitary $\ui$.)
It satisfies 
\begin{eqnarray}
\label{eq:ui2}
\ui \pi(\Ai)\ui^{\ast}&=&\pi\bigl(\Theta(\Ai)\bigr), \quad \ui\Omei=\Omei,\\
\label{eq:ui3}
\ui^{2}&=&\id,\ \ui^{\ast}=\ui.
\end{eqnarray}
Define 
\begin{eqnarray*}
\Hil&\equiv&(\otimes_{i\ge1}\Hili)\otimes \Hilo, \\
\Ome&\equiv&(\otimes_{i\ge1}\Omei)\otimes \Omeo, \\
\Wi&\equiv&
 (\uone \otimes \cdots \otimes u_{i-1} \otimes \id_{i} \otimes 
\cdots)\otimes 
\id_{0}, \ {\text{for}}\ (i\ge 2),\quad 
\Wf\equiv\id,\\
\Wzero&\equiv&(\otimes_{i\ge 1} \ui)\otimes \id_{0}, \\
\piib(\Ai)&\equiv&\bigl(\id_{1}\otimes \cdots \otimes \id_{i-1}
 \otimes \pii(\Ai)\otimes \id_{i+1}\otimes \cdots \bigr)\otimes \id_{0},\ 
 \Ai \in \AlIi,\ (i\ge 1), \\
\piob(\Ao)&\equiv&\bigl(\id_{1}\otimes  \id_{2}\cdots
 \bigr)\otimes \pio(\Ao),\ 
 \Ao \in \AlIo,
\end{eqnarray*}
where $\otimes_{i\ge 1}\Hili$ is an ordinary 
tensor product if the index set 
 is finite, while it is the incomplete infinite tensor product of
 $\Hili$, $i=1,2,\cdots,$ containing the vector 
$\otimesOmei$ (\cite{vonNeumann38})
 if the index set is infinite.
 Since $\ui\Omei=\Omei$, the infinite tensor product 
$\otimesui$ is well-defined in the latter case
 and it is a unitary leaving $\otimesOmei$ invariant.
Then there exists a unique representation $\pi$ 
 of $\AlI$ in $\Hil$ 
 satisfying
\begin{eqnarray}
\label{eq:pidef}
\pi(\ak)=\Wi\piib(\ak),\ \pi(\akcr)=\Wi\piib(\akcr) 
\end{eqnarray}
for $k \in \Ii$ and $i\ge 0$, because 
 $\pi(\ak)$ and $\pi(\akcr)$ 
 satisfy CAR.
The state 
\begin{eqnarray}
\label{eq:vpdef}
\vp(A)\equiv \bigl(\Ome,\,\pi(A) \Ome\bigr),\quad A \in \AlI
\end{eqnarray}
 gives a product state extension of $\vpiset$,
 as can  be immediately shown.

This proves the existence part of Theorem $\ref{thm:PRODUCT}$
 (1).
\begin{rem}
 In the above construction, the following formulae
 for $\Aipm\in \AlIipm$ hold.
\begin{eqnarray}
\pi(\Aip)=\piib(\Aip),\quad \pi(\Aim)=\Wi \piib(\Aim).
\end{eqnarray}
\end{rem}
\ \\
(b) {\underline{Necessity of the condition for $\vpi$.}}\\
We now show that all  states $\vpi$ except at most one must be
 even if a product state extension $\vp$ of $\vpiset$
 exists. 
Assuming that two states $\vpf$ and $\vps$ are not even,
 we show contradiction.
 Since $\vpf$ and $\vps$ are not even, there exists 
 $\Af\in \AlIfm$ and  $\As\in \AlIsm$ such that 
$\vpf(\Af)\ne0$, $\vps(\As)\ne0$. Then 
\begin{eqnarray}
\label{eq:contradict1}
\vp(\Af \As)=\vpf(\Af)\vps(\As)\ne0.
\end{eqnarray}
For $\Af \in \AlIfm$,
 both $\Af+\Af^{\ast}$ and  
$i(\Af-\Af^{\ast})$ are self-adjoint elements of $\AlIfm$.
Since $\Af$ is their linear combination,
 the value of $\vpf$ for one of them must be non-zero.
 Hence we may assume $\Af=\Af^{\ast}$, $\vp(\Af)\ne 0$
 for some $\Af \in \AlIfm$.
By the same reason, we may also assume that $\As=\As^{\ast}$,
 $\vp(\As)\ne 0$.
 Then both $\vp(\Af)$ and $\vp(\As)$
 are non-zero real numbers.
On the other hand,
 \begin{eqnarray*}
(\Af\As)^{\ast}=\As^{\ast}\Af^{\ast}=\As\Af=-\Af\As,
\end{eqnarray*}
the last equality being due to (\ref{eq:epsCOMMUT})
 where  $\sig=\sigp=-$ for the present case.
Therefore $\Af\As$ is skew self-adjoint and 
$\vp(\Af\As)$ must be pure imaginary.
We now have a pure imaginary 
$\vp(\Af\As)$ and non-zero real $\vp(\Af)\vp(\As)$
 which contradict with (\ref{eq:contradict1}).

\ \\
(c) {\underline{Uniqueness  of a product 
 state extension. }}\\
The linear span of  monomials
$\Af\cdots \Ak$ with $\mathcal{A}_l \in \mathcal{A}(\mathrm{I}_{i_l})$
 for all possible finite index sets 
$(\ione,\cdots,\ik)$ for all possible $k\in \NN$
 is dense in 
$\AlI$.
Therefore the values 
 on such monomials uniquely determine  $\vp$.

\ \\
(d) {\underline{Equivalence of evenness of  $\vp$  
 and that of  all $\vpi$ }\\
We show
 that the extension $\vp$ is even if and only if 
all  $\vpi$ are even.
  
If $\vp$ is even, then
$\vp\Theta=\vp$ implies $\vpi\Theta=\vpi$ by restriction.

In the converse direction, if all $\vpi$
 are even, then 
$\vp$
 satisfies 
\begin{eqnarray*}
\vp(A)=\vp(\Theta(A))
\end{eqnarray*} 
for all monomials
$A=\monomials$ by  the product property of $\vp$
 and  evenness of $\vpi$.
Thus
$\vp$ 
is even  by the same reasons as (c).

\subsection{Theorem \ref{thm:PRODUCT} (2)}
\label{subsec:PSE2}
(I) \underline{Case of a pair of 
pure states}\\
Assume that   states $\vpf$ and $\vps$ are both pure and that 
there exists their  joint extension $\vp$.
We show that  at least one of $\vpf$ and $\vps$ is  even
 and $\vp$ is their  product state extension.
 

By the assumption that $\vpf$ is pure, the representations 
$\pivpf$ and $\pivpft$ are irreducible and
 the following two cases cover all situations.

($\alpha$) $\pivpf$  and $\pivpft$ are unitarily equivalent.
  
($\beta$) $\pivpf$  and $\pivpft$ are disjoint.

We need the  following lemma for dealing with the case 
($\alpha$).
\begin{lem}
\label{lem:uone}
If $\vpf$ is pure and if $\pivpf$ and $\pivpft$ are unitarily 
 equivalent, there exists a self-adjoint  unitary 
 $\uone\in \pivpfAlIfpwpri$ satisfying 
(\ref{eq:uone1}) and  (\ref{eq:uone2}).
\end{lem} 
\proofs\ 
Since  $\pivpf$ and $\pivpft$ are assumed 
to be unitarily 
 equivalent, there exists a unitary $\uonehat$ on  $\Hilvpf$
 satisfying 
\begin{eqnarray*} 
\uonehat \pivpf(\Af) \uonehat^{\ast}=\pivpf(\Theta(\Af)),\quad \Af\in \AlIf.
\end{eqnarray*}
By $\Theta^{2}=\id$, $\uonehat^{2}$
 commutes with $\pivp(\AlIf)$.
 Since $\vpf$ is pure, $\pivpf$ is irreducible 
 and $\pivpfAlIfwpri=\BHvpf$.
 Hence $\uonehat^{2}=e^{i \theta}\id$.
  By setting $\uone\equiv \pm e^{-i\theta/2}\uonehat$,
 we have a self-adjoint unitary $\uone$ satisfying 
(\ref{eq:uone1}).
 We choose the sign $\pm$ so that (\ref{eq:uone2})
 is satisfied.

 Any $x\in \BHvpf$ has a decomposition
 $x=\xp+\xm$, $\xpm\equiv \frac{1}{2}(x\pm uxu)$.
 There exists a net $A_{\alpha}\in \AlIf$ such that 
 $\pivpf(A_{\alpha})\to x$ due to $\BHvpf=\pivpfAlIfwpri$.
Then $A_{\alpha\pm}\equiv 
\frac{1}{2}\bigl(A_{\alpha}\pm \Theta(A_{\alpha})  \bigr)\in \AlIfpm$
 and $A_{\alpha\pm}\to \xpm$.
  Hence $\xp\in \pivpfAlIfpwpri$.
 Since $\uone\uone\uone=\uone$ and $\uone=(\uone)_{+}$,
 $\uone  \in \pivpfAlIfpwpri$.
\proofend 

We resume the  proof for the case ($\alpha$).
Since $\pivpAlIsp$ commutes with $\pivpAlIf$ elementwise
 by (\ref{eq:epsCOMMUT}), 
Lemma \ref{lem:TAKESAKI4.11} implies 
 \begin{eqnarray}
\label{eq:plussplit}
\vp(\Af\Asp)=\vpf(\Af)\vps(\Asp)
\end{eqnarray}
 for all $\Af \in \AlIf$ and $\Asp \in \AlIsp$
 due to  the purity of $\vpf$.

We  want to  derive the same formula for $\Asm \in \AlIsm$.
 By Lemma \ref{lem:tech2} and the equivalence 
 of $\pivpf$ and $\pivpft$,
 the representations 
 $\pivp|_{\AlI}$ and $\pivpf$ are quasi-equivalent.
 Therefore, via the extension of the isomorphism
$\pivpf(A)\to \pivp(A)$, $A \in \AlIf$ to their weak closures,
 there exists  a self-adjoint unitary $\Uone\in \pivpAlIfpwpri$
 (corresponding to $\uone \in  \pivpfAlIfpwpri$ of Lemma
\ref{lem:uone})
 which satisfies 
 \begin{eqnarray} 
\label{eq:UONEEQ}
\Uone \pivp(\Af) \Uone=\pivp(\Theta(\Af))
\end{eqnarray}
 for all  $\Af\in \AlIf$.
 Since $\AlIfp$ commutes with $\AlIs$, 
we have $\Uone \in \pivpAlIspri$.
(We note that $\bar{\vp}(x)=\bar{\vpf}(x)$ for 
$x\in \pivpAlIfwpri\sim \pivpfAlIfwpri$ 
 by the identification of these von Neumann algebras 
via the isomorphism.)
 By (\ref{eq:epsCOMMUT}) and (\ref{eq:UONEEQ}),
 we have $\Uone \pivpAlIsm\in \pivpAlIfpri$.
  We now apply Lemma \ref{lem:TAKESAKI4.11}
 to  
\begin{eqnarray*}
\pivpAlIf \pivpAlIsm=
 \{\pivpAlIf \Uone\} \{\Uone \pivpAlIsm\}
\end{eqnarray*}
  to obtain 
\begin{eqnarray}
\label{eq:minussplit}
\vp(\Af \Asm)&=&\vpb(\pivpAf \pivpAsm)\nonum \\
&=&\vpb(\pivpAf \Uone) \vpb(\Uone\pivpAsm)
\end{eqnarray}
 for all $\Af \in \AlIf$ and 
$\Asm \in \AlIsm$,
 where we have used  purity of the  restriction 
 of $\vpb$ to $\pivpAlIfwpri$, 
 to which $\pivpAf\Uone$ belongs.

Since $\vps$ is pure, 
 we can  apply Lemma \ref{lem:TAKESAKI4.11}
 to  the pair $\pivpAs\in \pivpAlIs$
 and $\Uone \in \pivpAlIfpwpri \subset \pivpAlIspri$
 to obtain
\begin{eqnarray}
\label{eq:split37}
\vpb(\Uone \pivpAsm )=\vpb(\Uone)\vp(\Asm)=\vpb(\Uone)\vps(\Asm).
\end{eqnarray}
By setting $\Af=\id$ in (\ref{eq:minussplit})
 and substituting (\ref{eq:split37}) there,
 we obtain
\begin{eqnarray}
\label{eq:split38}
(1-\vpb(\Uone)^{2})\vps(\Asm )=0.
\end{eqnarray}
 We have two alternatives ($\alpha$1) and ($\alpha$2).\\
\quad ($\alpha$1) $\vps(\Asm )=0$ for all $\Asm\in \AlIsm$.\\
 In this case, $\vps$ is even. By (\ref{eq:minussplit})
 and (\ref{eq:split37}),
 \begin{eqnarray}
\label{eq:split39}
\vp(\Af \Asm)=0=\vpf(\Af)\vps(\Asm).
\end{eqnarray}
 By this and (\ref{eq:plussplit}), $\vp$ has the  product
 state property.
\ \\
\quad ($\alpha$2) $\vpb(\Uone )=\pm 1$.\\
 In this case, 
$\vpb(\Uone)=\overline{\vpf}(\uone)= 1$ by (\ref{eq:uone2}).
 Then
 \begin{eqnarray*}
\Vert \Omevp- \Uone \Omevp \Vert^{2}=
\Vert \Omevp \Vert^{2}+\Vert \Uone \Omevp \Vert^{2}
- 2 \vpb(\Uone)=0.
\end{eqnarray*}
Hence $\Uone\Omevp=\Omevp$
 and
 \begin{eqnarray*}
\vp(\Af)&=&(\Vecvp,\,\pivpAf \Vecvp)=
 (\Uone \Vecvp,\,\pivpAf \Uone \Vecvp)\nonum \\
&=& (\Vecvp,\,\Uone \pivpAf \Uone \Vecvp)=
 (\Vecvp,\,\pivp(\Theta(\Af) ) \Vecvp)=\vp(\Theta(\Af)),
\end{eqnarray*}
 so that $\vpf$ is even.
 Furthermore
 \begin{eqnarray*}
\vpb(x \Uone)&=&(\Vecvp,\, x \Uone \Vecvp)
=(\Vecvp,\,x \Vecvp)
=\vp(\Af)=\vpf(\Af),\nonum \\
\vpb(\Uone y)&=&
   (\Vecvp,\,\Uone y \Vecvp)
=(\Vecvp,\, y \Vecvp)
=\vp(\As)=\vps(\As)
\end{eqnarray*}
for $x=\pivp(\Af)$ and  $y=\pivp(\As)$.
Substituting them into (\ref{eq:minussplit}),
 we obtain  
\begin{eqnarray}
\label{eq:minussplit0}
\vp(\Af \Asm)=\vpf(\Af)\vps(\Asm).
\end{eqnarray}
Combing (\ref{eq:plussplit}) and (\ref{eq:minussplit0}),
$\vp$ is a product state extension of 
$\vpf$ and $\vps$.

We now deal with  the case ($\beta$).
By (\ref{eq:perpEQ}), 
\begin{eqnarray}
\label{eq:beta-minussplit1}
\vp(\Af \Asm)=(\pivp(\Af^{\ast})\Omevp,\, \pivpAsm\Omevp)=0,
\end{eqnarray}
 By setting $\Af=\id$, we obtain 
\begin{eqnarray}
\label{eq:beta-minussplit2}
\vp(\Asm)=0.
\end{eqnarray}
Thus (\ref{eq:split39}) holds and   $\vps$ is even.
By purity of $\vpf$, we have (\ref{eq:plussplit}) as in the case of (I). 
Therefore
$\vp$ is a product state extension of 
$\vpf$ and $\vps$.

For all cases, 
 uniqueness of the extension $\vp$
 follows from Theorem \ref{thm:PRODUCT} (1). 

\ \\
(II) {\underline{Purity  of the product  state extension}}\\
We prove 
that the (unique) product state  extension  $\vp$
 of $\vpiset$ is pure if all $\vpi$ are pure.
We use the notation of (a) in Subsection \ref{subsec:PSE1}.
 
 If $\vpi$ is pure, then 
$\piiAlIidoub=\BHi$ and $\ui\in \BHi$, 
which implies
\begin{eqnarray*}
\Wi\in \bigl(\otimes_{j=1}^{i-1}  \BHj \bigr) \otimes \bigr(\otimes_{j\ge i} \id_{j}\bigr)\otimes 
\id_{0}
\end{eqnarray*}
for $i\ge 1$. 
Starting with $i=1$, we obtain recursively
\begin{eqnarray*}
\pi\bigl(\Al(\cup_{j=1}^{i} \Ij)\bigr)^{\prime\prime}=
\bigl(\otimes_{j=1}^{i}  \BHj \bigr) \otimes \bigr(\otimes_{j\ge i+1} 
\id_{j}\bigr)\otimes 
\id_{0}
\end{eqnarray*}
 for $i\ge 1$.
 Hence 
\begin{eqnarray}
\pi\bigl(\Al(\cup_{j\ge 1} \Ij)\bigr)^{\prime\prime}&=&
\Bigl(\cup_{i \ge 1} \pi
\bigl(  
\Al(\cup_{j=1}^{i} \Ij)
\bigr)^{\prime\prime}
\Bigr)^{\prime\prime}\\
&=&\bigl(\otimes_{j\ge 1} \BHj \bigr) \otimes \id_{0}.
\end{eqnarray}
 Since $\Wzero$ belongs to this algebra,
 we obtain 
 \begin{eqnarray*}
\pi\bigl(\Al(\cup_{j\ge 0} \Ij)\bigr)^{\prime\prime}
=\bigl(\otimes_{j\ge 1} \BHj \bigr) \otimes \BHo=\BH.
\end{eqnarray*}
Hence $\vp$ is pure.

\ \\
(III) {\underline{General case}}\\
We first prove 
that 
 if a state of one subsystem, say $\vpf$, is 
 not even, then 
$\vpi$ for $i \ge 2$ have to be all even.
 We take an individual $i \ge 2$
 and apply the result proved for the case (I)
 to the pair of states $\vpf$ and $\vpi$
 of two subsystems $\AlIf$ and $\AlIi$,
 which are pure and for which the restriction 
of $\vp$ to $\Al(\If\cup \Ii)$ is a joint 
 extension, thus obtaining the desired conclusion that 
 $\vpi$ must be even because $\vpf$ is not even.

Next
 we use both (I) and (II) to prove the product 
 property (\ref{eq:EXTk}) for $\vp$.
Let $\Ikup\equiv\cup_{i=1}^{k}\Ik$
 and $\vpkup$ be the restriction of 
 $\vp$ to $\Al(\Ikup)$.
For $k=2$, $\vpsup$ is a joint extension of 
 pure states $\vpf$ and $\vps$.
 By (I), $\vpsup$ is a product state extension of 
$\vpf$ and $\vps$.  By (II), $\vpsup$ is pure.
 Inductively, assume that 
 $\vpkmoup$ is a product state extension 
 of $\vpf,\cdots,\vp_{k-1}$ and a pure state.
 Then $\vpkup$ is a joint extension of 
 $\vpkmoup$ and $\vpk$ and hence
 satisfies (\ref{eq:EXTk}).
It is pure by (II).
\proofend

\subsection{Theorem \ref{thm:PROEXT-1} (1)}
\label{subsec:PROEXT-1-1}
 We show that $\vpf$ is pure if 
  $\vp$ is pure, by showing that $\vp$
   is not pure if $\vpf$ is not pure.

For non-pure $\vpf$,
 there exist two distinct states
 $\vpfal$ and $\vpfbe$ of $\AlIf$
 such that 
$\vpf=\lam\vpfal+(1-\lam) \vpfbe$ for some 
 $\lam\in (0,\,1)$.%

The restriction $\psi$ of $\vp$ to 
$\Al(\cup_{i \ge 2}\Ii)$ is even,
  being the product state extension 
   of even $\vpi$, $i\ge 2$.
Hence there  exist product state 
 extensions $\vpal$  of the pair $\vpfal$ and $\psi$
  and $\vpbe$
 of the pair  $\vpfbe$ and $\psi$ due to Theroem
 \ref{thm:PRODUCT} (1):
\begin{eqnarray*}
\vpal(\Af\As)=\vpfal(\Af)\psi(\As),\ 
\vpbe(\Af\As)=\vpfbe(\Af)\psi(\As)
\end{eqnarray*}
 for $\Af \in \AlIf$ and  $\As \in \Al(\Isup)$.
 Hence $\vp$ and $\lam\vpal+(1-\lam) \vpbeta$
 have the same value for $\Af\As$.
 Since the linear span of such $\Af \As$ is dense
 in $\Al(\Ifup)$, we obtain
 $\vp=\lam\vpal+(1-\lam) \vpbeta$.
 Since $\vpfal\ne \vpfbe$,  we have $\vpal\ne \vpbe$.
 Therefore $\vp$ is not pure.
%
%

\subsection{Theorem \ref{thm:PROEXT-1} (2)}
\label{subsec:PROEXT-1-2}

We already know that $\vp$ is pure 
 if all $\vpi$ are pure by Theorem \ref{thm:PRODUCT} (2).
We now prove 
the only if part. 
We first consider the case 
  where $\vp$ is a product state extension 
   of $\vpf$ and $\vps$ for two subsystems.
    We are in the situation where $\vp$ is pure,  
    $\vps$ is even, $\vpf$ is pure 
    (by Theorem \ref{thm:PROEXT-1} (1)),
   $\pivpf$ and $\pivpft$
 are not disjoint, and hence are unitarily equivalent 
  (the case ($\alpha$) of \ref{subsec:PSE2}
(I)); 
  therefore, we have a self-adjoint unitary $\Uone \in \pivpAlIfpwpri$
  satisfying (\ref{eq:UONEEQ}).
Now our aim is to show  purity of $\vps$.

We define
\begin{eqnarray} 
\label{eq:pis0}
\pis(\As)=\pis(\Asp+\Asm)\equiv \pivp(\Asp)+\Uone \pivp(\Asm)
\end{eqnarray}
 for $\As=\Asp+\Asm$, $\Aspm\in \AlIspm$.
Since $\Uone^{\ast}=\Uone$, $\Uone^{2}=\id$ and 
$\Uone\in \pivpAlIfpwpri$ commutes with $\pivpAlIs$,
 $\pis$ is a representation
 of $\AlIs$, commuting with $\pivpf$
  due to (\ref{eq:epsCOMMUT}) and (\ref{eq:UONEEQ})
and $\pivpAlIf \cup \pis(\AlIs)$ genetates 
$\bigl\{ \pivpAlIf \cup \pivpAlIs\ \}^{\prime\prime}$,
 which is $\BHvp$ 
  due to purity of $\vp$.
Therefore 
 $\vpb$
 is  a product state of mutually commuting 
 $\pivpAlIfwpri$
 and $\pisAlIswpri$
  by Lemma \ref{lem:TAKESAKI4.11} (due to 
  purity of $\vpf$)  and 
\begin{eqnarray} 
\label{eq:vpstil0}
\vpstil(\As)\equiv(\Vecvp,\,\pis(\As)\Vecvp),\quad \As \in \AlIs
\end{eqnarray}
is a pure state of $\AlIs$.
By the product state property of $\vp$
 and evenness of $\vps$, we have
\begin{eqnarray*} 
(\Vecvp,\,\pis(\Asm) \Vecvp)=
(\Vecvp,\,\Uone\pivpAsm \Vecvp)=(\Vecvp,\,\Uone\Vecvp)\vps(\Asm)=0
\end{eqnarray*}
 for $\Asm\in \AlIsm$\,. 
 Hence 
\begin{eqnarray*} 
\vpstil(\As)=\vps(\As),\quad \As \in \AlIs.
\end{eqnarray*}
Thus $\vps=\vpstil$, and hence $\vps$ is pure, which is 
our desired result.

For the general case of many subsystems, we use 
this result for two subsystems and 
 Theorem \ref{thm:PROEXT-1} (1) inductively
 to obtain  purity of 
 both the restriction 
of $\vp$ to 
$\Al(\cup_{j \ge i}\Ij)$
 and $\vpi$
 for $i=2,3,\cdots$.

The case of even $\vpf$ is a special case of unitarily equivalent
 $\pivpf$ and $\pivpft$.
\proofend 

\subsection{Theorem \ref{thm:PROEXT-D} (1)}
\label{subsec:PROEXT-D2}

Assume that $\vp$ is a product state extension 
 of $\vpf$ and even $\vps$ and that 
 $\pivpf$ and $\pivpft$
 are disjoint.
First we assume purity of  $\vp$
 and prove the only if part.
By Theorem \ref{thm:PROEXT-1} (1),
$\vpf$ is pure.
We have only to prove that
 $\vpsp$ is pure.  

 Let $\DAl\equiv \pivpAlIf\cup \pivpAlIsp$.
By (\ref{eq:perpEQ}),
\begin{eqnarray*}
\Hilvp=\Hilvpp\oplus  \Hilvpm,
\end{eqnarray*}
 and $\DAl$
 leaves $\Hilvppm$ invariant.

Let $v\equiv \ai+\aicr$ for 
 a fixed  $i$ belonging to  $\Is$.
 Set 
 \begin{eqnarray}
\label{eq:Vs}
\Vs\equiv \pivp(v).
 \end{eqnarray}
Then $v$ and $\Vs$ are self-adjoint unitaries
 and 
\begin{eqnarray}
\label{eq:ipureD7}
\pivp\bigl(\AlIs \bigr)&=&\pivpAlIsp+ \pivpAlIsp\Vs,\\
\label{eq:ipureD8}
\Vs\Hilvppm&=&\Hilvpmp,\\
\label{eq:ipureD9}
\Vs \pivpAf \Vs&=&\pivpAft,\quad \Af \in \AlIf.
\end{eqnarray}

We now show that the restriction
 of $\DAl$ 
 to $\Hilvpp$ is irreducible.
 Let  
\begin{eqnarray*}
\DMlpm\equiv  \DAl^{\wpri} \cap \BHvppm,
\end{eqnarray*}
 where $\BHvppm$ are  imbedded in $\BHvp$.
Take $x \in (\DMlp)^{\prime} \cap \BHvpp$.
Then $\Vs x\Vs \in \BHvpm$ due to (\ref{eq:ipureD8}).
For any
$y\in \DMlm$, we have $\Vs y \Vs\in \DMlp$ also
 due to (\ref{eq:ipureD8})
 and 
\begin{eqnarray*}
(\Vs x \Vs)y=\Vs x (\Vs y \Vs)\Vs=\Vs (\Vs y \Vs) x\Vs=y(\Vs x \Vs).
\end{eqnarray*}
Hence $\Vs x \Vs\in \DMl^{\prime}_{-}$
 and 
 $x\oplus \Vs x \Vs$ is in $\DAl^{\prime}$.
Since   $x\oplus \Vs x \Vs$ commutes with $\Vs$
 due to (\ref{eq:ipureD8}), it belongs to
$\Bigl(\pivpAlIf \cup \pivpAlIs\Bigr)^{\prime}=\BHvp^{\pri}$
 due to purity of $\vp$.
Therefore $x\oplus \Vs x \Vs$
 is a multiple of identity and so is 
 $x$.
This shows that 
$\DMlp=\BHvpp$ is irreducible.
Then  $\vpb$ restricted to $\DMlp$ is 
 a product state of the commuting pair 
 $\pivpAlIfwpri$ and $\pivpAlIspwpri$
 which generate $\BHvpp$.
 Hence its restriction to $\pivpAlIspwpri$
  is pure and $\vpsp$ is pure.

To prove the converse, 
 assume that $\vpf$
 and $\vpsp$ are  pure and that  
$\pivpf$ and $\pivpft$ are disjoint.
Let $\vp$ be  the product  state extension of $\vpf$ and $\vps$.
 We shall show  purity of $\vp$.

Since $\pivpAlIf$ restricted to $\Hilvpp$
 and to $\Hilvpm$ do not have any non-zero intertwiner,
 $x \in \pivpAlIfuspri$
 has to be of the form $x=x_{+}\oplus  x_{-}$, where $x_{\pm}\in \BHvppm$.
 Since $\vp$ is a product state of $\vpf$ and $\vpsp$
 (of a commuting pair $\AlI$ and $\AlIsp$),
 both of which are pure,
$\DAl^{\wpri}$ is irreducible
 on $\Hilvp=\overline{\DAl\Omevp}$
  and hence 
$x_{+}=c \id_{\Hilvpp}$.
Since $\Vs$ commutes with $x$, we have 
$x_{-}=\Vs x_{+}\Vs=c {\id}_{\Hilvpm}$.
Therefore, we obtain $x=c \id$.
Hence $\pivpAlIfus$ is irreducible and 
 $\vp$ is pure.
%
\subsection{ Theorem \ref{thm:PROEXT-D} (2)}
\label{subsec:PROEXT-D2}
The if part is evident.
 We  prove the only if part.
 Let  
\begin{eqnarray*}
\label{eq:Hil}
\Hil&\equiv&\Hilvpsp \oplus \Hilvpsp,\\
\label{eq:Vec}
\Vec&\equiv&\Vecvpsp\oplus  0,  \\
\label{eq:K}
K (\xip\oplus  \xim)
&\equiv& \xim\oplus \xip,\quad \xip\oplus  \xim\in \Hil,\\
\label{eq:pis1}
\pis(\Asp)&\equiv&\pivpsp(\Asp)\oplus \pivpsp(v \Asp v),\ 
{\text{for}}\ \Asp \in \AlIsp \\
\label{eq:pis2}
\pis(\Asp+ \Asp^{\prime}v)
&\equiv&\pis(\Asp)+ \pis( \Asp^{\prime} )K,\ 
{\text{for}}\ \Asp,\Asppri \in \AlIsp,
\end{eqnarray*}
where   $v\in \AlIsm$ is the same as in (\ref{eq:Vs})
 and $(\Hilvpsp$, $\pivpsp$, $\Vecvpsp)$ 
 is the GNS triplet for $\vpsp$.
 
 By a straightforward computation, 
 we see that $\pis$ is a representation 
of $\AlIs$,
and by evenness of $\vps$,
\begin{eqnarray}
\label{eq:2ipureD}
(\Vec,\,\pis(\As)\Vec)=\vps(\As),\quad \As\in \AlIs.
\end{eqnarray}
 Since $\Vec$ is  cyclic for $\pis(\AlIs)$,
 $(\Hil,\,\pis,\,\Vec)$ is unitarily 
 equivalent to  $(\Hilvps,\,\pivps,\,\Vecvps)$
 and purity of $\vps$ is equivalent to  irreducibility 
of $\pis$.

By identifying $\Bl(\Hil)$ with 
${\mathrm{M}}_{2}(\BHvpsp)$, we have
\begin{eqnarray}
\label{eq:ipureD26}
\pis(\Asp+ \Asp^{\prime}v)
=\left( 
\begin{array}{cc}
\pivpsp(\Asp) & \pivpsp(\Asppri) \\
\pivpsp(v \Asppri v)  & \pivpsp(v \Asp v)
\end{array}
\right).
\end{eqnarray}
To determine  $\pisAlIspri$, 
let
\begin{eqnarray}
\label{eq:ipureD27}
\left( 
\begin{array}{cc}
a & b\\
c  & d 
\end{array}
\right) \in  \pis(\AlIs)^{\prime}.
\end{eqnarray}
We see that  
the commutativity of (\ref{eq:ipureD26})  and 
 (\ref{eq:ipureD27}) is equivalent to the 
 condition that 
$a=d$ is a scalar and $b=c$ satisfies
\begin{eqnarray}
\label{eq:Dlist5}
\pivpsp(v\Asp v)c=c\pivpsp(\Asp).
\end{eqnarray}

 From (\ref{eq:Dlist5}), we see that $c^{2} \in \pivpspAlIsppri$
 due to $v^{2}=\id$.
 Therfore $c^{2}$ is a scalar.
 If $c$ is not $0$, we may assume  $c^{2}=\id$
 (by  multiplication of a scalar).
 If $c_1$ and $c_2$ satisfy (\ref{eq:Dlist5}),
 then $c_1 c_2$ must be a scalar by the same reason.
 Hence $c$ is unique up to a multiplication 
 of a scalar.
 If   $c$ satisfies  (\ref{eq:Dlist5}), 
$c^{\ast}$ also satisfies (\ref{eq:Dlist5}).
 By  uniqueness $c^{\ast}=e^{i \theta}c$  for some 
 $\theta \in \R$.
Since  $c^{\ast}c=e^{i \theta}c^{2}=e^{i \theta}$
 must be a positive real, we have $c^{\ast}c=\id$
 and hence $c=c^{\ast}$.
 Namely $c$ is a self-adjoint unitary.
For $\vps$ to be not pure,
 $\pisAlIspri$ has to be non-trivial
  and hence such a $c$ exists.

Since  $c$ belongs to  $\pivpspAlIspwpri=\BHvpsp$, 
\begin{eqnarray}
\label{eq:pishat}
\pishat(\Asp+ \Asppri v)
\equiv \pivpsp(\Asp)+ \pivpsp(\Asppri) c
\end{eqnarray}
  is  a
   representation of $\AlIs$ on $\Hilvpsp$ due to 
(\ref{eq:Dlist5}).
 Since $\pivpsAlIsp$ is already irreducible on $\Hilvpsp$,
 so is $\pisd(\AlIs)$ and 
\begin{eqnarray}
\label{eq:vpshat}
\vpshat(A)
\equiv (\Vecvpsp,\,\pishat(\As) \Vecvpsp)
\end{eqnarray}
 is a pure state  of $\AlIs$.

Since $\vpshat$ and $\vps$ coincide on $\AlIsp$,
 we have 
 \begin{eqnarray*}
\frac{1}{2}(\vpshat+\vpshat \Theta )=\vps
\end{eqnarray*}
due to evenness of $\vps$.

To prove that $\pi_{\vpshat}$ and $\pi_{\vpshatt}$
 are disjoint, assume the contrary.
 
 Since $\pi_{\vpshat}$ is irreducible,
  $\pi_{\vpshat}$ and $\pi_{\vpshatt}$ 
   are unitarily equiavlent 
    and   there exists a unitary $\utwo$ 
 on $\Hilvpshat$ implementing $\Theta$ on the representation 
 $\pi_{\vpshat}$ by Lemma \ref{lem:uone}.

 However, it has  to commute with $\pi_{\vpshat}(\AlIsp)$,
 which is irreducible on $\Hilvpshat$
 by Theorem \ref{thm:PROEXT-D} (1).
Hence $\utwo$ has to be trivial and cannot implemented 
a non-trivial automorphism $\Theta$.
 Hence  $\pi_{\vpshat}$ and $\pi_{\vpshatt}$
 are disjoint. 
 \proofend
\section{Pair of  General States--Theorem \ref{thm:F-NONPURE}}
\label{sec:F-NONPURE}

It is convenient to prove Theorem \ref{thm:F-NONPURE}
at this stage. 
Let $\vp$ be a joint extension of $\vpf$ and $\vps$.
Assume ($\beta$) in $\S$ \ref{subsec:PSE2},
namely,  assume that $\pivpf$ and $\pivpft$
 are disjoint.
 This  implies (\ref{eq:beta-minussplit1}) 
 and (\ref{eq:beta-minussplit2}).
Hence $\vps$
 is even.
Conversely, if $\vps$ is even, 
 then a product state extension of $\vpf$
 and $\vps$ exists by Theorem \ref{thm:PRODUCT} (1).
\proofend
\section{Pair of Pure and General States}
\label{sec:PUREGENERAL}
We prove Theorem \ref{thm:FPURE}
for a joint extension $\vp$ of $\vpf$
 and $\vps$ assuming purity of $\vpf$.
\subsection{Theorem \ref{thm:FPURE} (3)}
\label{subsec:FPURE3}
If $\pvpf=0$, (\ref{eq:EXTok})
 is equivalent to $\lamvps=1$
 since $1\ge \lamvps$.
If $\vps$ is even, then $\vps=\vpst$
and $\lamvps=1$.
 Conversely, if $\lamvps=1$,
 then $\psi\equiv\vps-\vpst\ge 0$
 by (\ref{eq:lamineEQ}).
 Since $\psi(\id)=\vpf(\id)-\vpft(\id)=0$,
 we obtain $\psi=0$
 and hence $\vps$ is even.
\subsection{Theorem \ref{thm:FPURE} for the case ($\beta$)}
\label{subsec:beta}
We consider two alternative cases 
$(\alpha)$ ($\pivpf$ and $\pivpft$ 
are unitarily equivalent) and 
$(\beta)$ (they are mutually disjoint)
 separately.
For the case ($\beta$), 
 any representative vectors
$\phivec$ and $\psivec$ of $\vpf$
 and $\vpft$  have to be mutually orthogonal
 and hence $\pvpf=0$.
 
If (\ref{eq:EXTok}) holds, 
$\vps$ is even by 
$\S$ \ref{subsec:FPURE3} and the product state extension
 of $\vpf$ and $\vps$ exists.

Conversly, if  $\vp$ exists,
 $\vps$ is even and (\ref{eq:beta-minussplit1})
 holds  by the proof of $\S$ \ref{sec:F-NONPURE}
  and 
 (\ref{eq:plussplit})  holds by  purity of $\vpf$.  
Hence $\vp$ is the product state extension and 
(\ref{eq:EXTok})
  holds by evenness of $\vps$.

This also proves 
the necessity of (4-i) in (4).

We have established those parts of Theorem \ref{thm:FPURE}
 related to the case ($\beta$).
\subsection{Theorem \ref{thm:FPURE} (1) for the case ($\alpha$)}
\label{subsec:alpha}
We first show  the following lemma.
\begin{lem}
\label{lem:Trans}
The following formula holds under the assumption
 of Lemma \ref{lem:uone} and for $\uone$
 given by that Lemma.

 \begin{eqnarray}
\label{eq:Trans5}
\pvpf=\vpfb(\uone).
\end{eqnarray}
\end{lem}
\proofs\ 
In the situation of Lemma \ref{lem:uone},
 the transition probability between 
 the vector states $\vpfb$ by $\Vecvpf$
 and $\overline{\vpft}$ by $\uone \Vecvpf$
 for the algebra $\BHvpf=\pivpfAlIfwpri$ is shown 
 in \cite{UHLMANN77} to be given by 
\begin{eqnarray*}
\label{eq:Trans2}
P(\vpfb,\, \overline{\vpft})\equiv
 | (\Vecvpf,\,\uone \Vecvpf) |^{2}
\end{eqnarray*}
Due to (\ref{eq:uone2}), we obtain  (\ref{eq:Trans5}).
\proofend

We resume the proof of  Theorem \ref{thm:FPURE} for the case 
($\alpha$).\\
(a) {\underline{Necessity of (\ref{eq:EXTok}) for (1)}}\\
 The equations (\ref{eq:plussplit})
 and (\ref{eq:minussplit}) hold by purity of $\vpf$.
 By setting $\Af=\id$ in  (\ref{eq:minussplit}),
\begin{eqnarray}
\label{eq:split11}
\vps(\Asm)&=&\vp(\Asm)=
\pvpf
\vpb(\Uone \pivpAsm)
\end{eqnarray}
due to (\ref{eq:Trans5})
 and $\vpb(\Uone)=\vpfb(\uone)$.
We consider the representation 
$\pis$ of $\AlIs$ given by (\ref{eq:pis0}). 
Then $\vpstil$ given by (\ref{eq:vpstil0})
 satisfies 
\begin{eqnarray}
\label{eq:vpstil5}
\vps+\vpst=
\vpstil+\vpstilt,\ \ \vps-\vpst=\pvpf(\vpstil-\vpstilt).
\end{eqnarray}
Hence
\begin{eqnarray*}
2 \pvpf\vpstil=(1+\pvpf)\vps-(1-\pvpf)\vpst\ge 0.
\end{eqnarray*}
Therefore (\ref{eq:EXTok}) holds because
\begin{eqnarray*}
\vps\ge \frac{1-\pvpf}{1+\pvpf} \vpst.
\end{eqnarray*}
\ \\
(b) {\underline{Sufficiency  of (\ref{eq:EXTok}) for (1)}}\\
If $\pvpf=0$, then (\ref{eq:EXTok})
 implies that $\vps$ is even.
  Hence the product state extension of $\vpf$
   and $\vps$ exists.
   
Assume  $\pvpf\ne0$ and set 
 \begin{eqnarray}
\label{eq:EXTform3}
\vpspri(\Asp+\Asm)
\equiv \vps(\Asp)+\frac{1}{\pvpf}
\vps(\Asm)
\end{eqnarray}
for $\Aspm \in \AlIspm$.
Then 
 \begin{eqnarray}
\label{eq:EXTform4}
\vpspri
&=&\frac{1}{2 \pvpf}\Bigl\{
\bigl\{1+\pvpf\bigr\}\vps-\bigl\{1-\pvpf\bigr\}\vpst
\Bigr)\ge 0
\end{eqnarray}
by (\ref{eq:EXTok}).
Hence $\vpspri$ is a state of $\AlIs$
 due to $\vpspri(\id)=\vps(\id)=1$.
Let
\begin{eqnarray}
\label{eq:suff4}
\Hil&\equiv& \Hilvpf\otimes \Hilvpspri,\quad 
\Vec\equiv \Vecvpf\otimes \Vecvpspri, \\
\label{eq:suff5}
\pi(\Af\As)&\equiv& \pivpf(\Af)\otimes \pivpspri(\Asp)+
\pivpf(\Af)\uone \otimes \pivpspri(\Asm),
\end{eqnarray}
for $\Af\in \AlIf$, $\As=\Asp+\Asm$, 
$\Aspm \in \AlIspm$.
Then operators  $\pi(\Af\As)$
 satisfy CAR and hence $\pi$ extends to a representation
 of $\AlIfus$.
 The state
\begin{eqnarray}
\label{eq:defvp}
\vp(A)\equiv(\Vec,\,\pi(A)\Vec),\quad A \in \AlIfus
\end{eqnarray}
of $\AlIfus$
 satisfies 
\begin{eqnarray*}
\vp(\Af)&=&\vpf(\Af), \nonum\\
\vp(\As)&=&\vps(\Asp)+\vpfb(\uone)\vpspri(\Asm)=\vps(\As),
\end{eqnarray*}
where (\ref{eq:Trans5}) is used in the last eqaulity.
Hence $\vp$ is a joint extension of 
 $\vpf$ and $\vps$.

\ \\
(c) {\underline{Proof of  (2)}}\\
By (a), $\vp$
satisfis (\ref{eq:plussplit})
 and (\ref{eq:minussplit}).
  Since (\ref{eq:split11})
   implies 
 \begin{eqnarray}
\label{eq:split13}
\vpb(\Uone\pivp(\Asm))&=&\frac{1}{\pvpf}\vps(\Asm),
\end{eqnarray}
(\ref{eq:plussplit})
 and (\ref{eq:minussplit})
imply 
(\ref{eq:EXTform1}) and (\ref{eq:EXTform2}).

\ \\
(d) {\underline{Necessity of (4-ii)}}\\
Assume $\pvpf=0$. (Then $\vps$ is even and the product state 
 extension exists if (\ref{eq:EXTok}) holds.)
  Assume the existence of a non-product 
   joint  extension $\vp$ of $\vpf$ and $\vps$.
  Then we are in the case ($\alpha$)
   by $\S$ \ref{subsec:beta}, namely (4-i) holds.
   Hence we may use  
    $\Uone\in \pivpAlIfpwpri$
 satisfying  (\ref{eq:UONEEQ}).
Let $\pis$ and $\vpstil$
 be given by (\ref{eq:pis0}) and  (\ref{eq:vpstil0}).
 Then $\pisAlIs$ commutes with $\pivpAlIf$
 and  $\vpstil$ is a state of $\AlIs$. 
  
  By purity of $\vpf$, 
 (\ref{eq:plussplit})
 and (\ref{eq:minussplit})
  hold,  implying 
  (\ref{eq:vppriform}).
  If $\vpstil=\vps$, then 
  $\vpstil(\Asm)=\vps(\Asm)=0$ 
   and $\vp$ is the product state extension.
    Hence $\vpstil \ne\vps$
in order that $\vp$
 is not the product state extension.
  On $\AlIsp$, $\vpstil$
   coincides with $\vps$.
    Therefore 
\begin{eqnarray}
\vpstil \ne \vpstil \Theta\ \ \text{and}\ \ 
\vps=\frac{1}{2}(\vpstil+ \vpstil\Theta).
\end{eqnarray}
\ \\
(e) {\underline{Proof of (5) and sufficiency of (4-ii)}}\\
By (d), any joint extension $\vp$
 has to be of the form   (\ref{eq:vppriform}).
 We show the exsistence of the GNS triplet for $\vp$
  given by   (\ref{eq:vppriform})
   and prove sufficiency of (4-ii) as well as (5).

Let
\begin{eqnarray}
\Hil&\equiv& \Hilvpf \otimes \Hilvpstil,\quad 
\Vec\equiv \Vecvpf \otimes \Vecvpstil,\\
\label{eq:vppriformrep}
\pipri(\Af \As)&\equiv& \pivpf(\Af)\otimes \pivpstil(\Asp)+
 \pivpf(\Af)\uone \otimes \pivpstil(\Asm),
\end{eqnarray}
for $\Af\in \AlIf$  and $\As=\Asp+\Asm$,
 $\Aspm \in \AlIspm$.
 Then operators $\pipri(\Af \As)$ satisfy CAR for $\AlIfus$
  and hence $\pipri$ extends to a representation of 
  $\AlIfus$.
  We have 
  \begin{eqnarray*}
(\Vec,\,\pipri(\Af\As)\Vec)&=&
\vpf(\Af)\vpstil(\Asp)+
 \vpfb(\pivpf(\Af)\uone)\vpstil(\Asm)\\
 &=&
 \vp(\Af\As).
\end{eqnarray*}
\proofend

\end{document}